\documentclass[11pt]{article}
\pdfoutput=1

\usepackage{standalone}

\graphicspath{{figures/}}
\usepackage{ulem}

\usepackage{euscript}
\usepackage{amsfonts}
\usepackage{amsbsy}
\usepackage{epsfig}
\usepackage{amsthm}
\usepackage{amscd}
\usepackage{amstext}
\usepackage{amsfonts}
\usepackage{verbatim}
\usepackage{cancel}
\usepackage{capt-of}
\usepackage{empheq}

\usepackage[nosort]{cite}
\usepackage{bm}
\usepackage{authblk}
\usepackage{esint}
\usepackage[T1]{fontenc}
\usepackage{mathdots}
\usepackage[centertableaux]{ytableau}

\usepackage[utf8]{inputenc}
\usepackage{graphicx}
\usepackage{physics}
\usepackage{mathtools}
\usepackage{bbold}

\usepackage{setspace}
\usepackage{colortbl}
\usepackage{xcolor}

\usepackage{amsmath}
\numberwithin{equation}{section}
\makeatletter
\renewcommand*\env@matrix[1][\arraystretch]{%
  \edef\arraystretch{#1}%
  \hskip -\arraycolsep
  \let\@ifnextchar\new@ifnextchar
  \array{*\c@MaxMatrixCols c}}
\makeatother
\usepackage{amssymb}
\usepackage{mathrsfs}
\usepackage{mathtools}
\usepackage{physics}
\usepackage{booktabs}
\usepackage{bbm}
\usepackage{float}

\usepackage{pgfplots}
\pgfplotsset{compat=1.15}
\usepackage{tikz}
\usetikzlibrary{external}
\usetikzlibrary{arrows}

\usepackage{subcaption}
\usepackage{graphicx}
\usepackage{mathtools}
\usepackage[many]{tcolorbox}
\tcbset{shield externalize}
\usepackage{xcolor, colortbl}
\usepackage[a4paper]{geometry}
\usepackage{amsthm}
\theoremstyle{definition}

\usepackage[hidelinks,linktocpage]{hyperref}
\hypersetup{
    colorlinks=true,
    linkcolor=blue,
    citecolor=blue,
    }

\usepackage{mathrsfs}

\renewcommand{\arraystretch}{2}
\def\ben{\begin{equation}}
\def\een{\end{equation}}
\def\half{{\textstyle{\frac{1}{2}}}}
\let\a=\alpha \let\b=\beta   
   \let\k=\kappa
 \let\m=\mu

\let\w=\omega

\let\pa=\partial
\def\be{\begin{equation}}
\def\ee{\end{equation}}
\def\beq{\begin{equation}}
\def\eeq{\end{equation}}
\def\ba{\begin{array}}
\def\ea{\end{array}}

\def\dalemb#1#2{{\vbox{\hrule height .#2pt
       \hbox{\vrule width.#2pt height#1pt \kern#1pt
               \vrule width.#2pt}
       \hrule height.#2pt}}}

\newcommand{\bea}{\begin{eqnarray}}
\newcommand{\eea}{\end{eqnarray}}

\def\vep{{\varepsilon}}

\makeatletter
\newcommand*\bigcdot{\mathpalette\bigcdot@{.5}}
\newcommand*\bigcdot@[2]{\mathbin{\vcenter{\hbox{\scalebox{#2}{$\m@th#1\bullet$}}}}}
\makeatother

\def\R{{{\mathbb R}}}

\def\Z{{{\mathbb Z}}}

\def\ocal{{\mathcal{O}}}

\title{A veiled BKL singularity and its holographic probes}
\author{Sean A. Hartnoll$^{\flat}$ and Alexander Zhiboedov$^\sharp$}
\affil{\it $^{\flat}$ Department of Applied Mathematics and Theoretical Physics, \\
\it University of Cambridge, Cambridge CB3 0WA, UK
}
\affil{\it $^{\sharp}$
CERN, Theoretical Physics Department, Geneva, Switzerland
}

\date{}

\begin{document}
\AddToHookNext{shipout/foreground}{%
  \put(0,-45){\makebox[\dimexpr\paperwidth-2.5cm\relax][r]{%
    \normalfont\small CERN-TH-2026-316}}%
}

\maketitle

\begin{abstract}
We analyse an M-theory realisation of BKL chaos in the interior of a black hole dual to nonzero-temperature ABJM theory. We show that BKL bounces in the interior lead to complex extremal surfaces which we call holographic veils. Holographic veils effectively hide the physical, real black hole interior from boundary two-point correlation functions, which instead are governed by a complex Reissner-Nordstr\"om-like singularity. We demonstrate, however, that the real black hole interior can be probed by analytically continued entanglement entropy and by a model for the one-point function of heavy operators. In this way we are able to read off the locations of veils, bounces and interior Kasner exponents.
\end{abstract}

\newpage

\tableofcontents

\newpage

\section{Introduction}
\label{sec:introduction}

Space and time exchange roles across a black hole horizon. Within the AdS/CFT correspondence this fact results in a remarkable relationship between thermal physics and cosmology \cite{Maldacena:2001kr}. The thermal state of holographic conformal field theories (CFTs) is described by the stationary, equilibrium exterior of black holes \cite{Witten:1998zw}. Yet the Lorentzian bulk geometries also contain a time-dependent cosmology in the black hole interior. The true holographic raison d'\^etre of this `bonus cosmology' remains only partially understood \cite{Maldacena:2013xja}. Meanwhile, the embedding of cosmology within AdS/CFT allows a well-defined holographic vantage point on the longstanding question of the fate of cosmological singularities in quantum gravity. This last research program was pioneered over twenty years ago \cite{Fidkowski:2003nf, Festuccia:2005pi}, but has seen considerable progress in the past few months \cite{Komargodski:2026oob, Batra:2026gbq, Buric:2026sit, Dodelson:2026hjm, Liu:2026luf}.

The classical approach to a cosmological singularity is known to be extremely rich \cite{Belinsky:1970ew, Misner:1969hg,Belinski:2017fas}. The chaotic evolution of the spacetime is commonly referred to as `cosmological billiards' \cite{Damour:2002et} or as `BKL dynamics', in reference to the seminal paper by
Belinsky, Khalatnikov, and Lifshitz \cite{Belinsky:1970ew}. It is a nontrivial and fascinating fact that this behaviour, discovered in four-dimensional vacuum gravity, extends to cosmological singularities in UV-complete theories including ten-dimensional string theory and eleven-dimensional M-theory \cite{Damour:2000wm,Damour:2000hv}. Our best-formulated version of nonperturbative string theory remains the AdS/CFT correspondence \cite{Maldacena:1997re}. The natural question arising is then: what is the holographic meaning of BKL chaos and what is its fate in the full quantum string theory?
In the present paper we will realise and probe BKL chaos within the M-theory/ABJM holographic correspondence \cite{Aharony:2008ug,Aharony:2008gk}.

The workhorse for holographic probes of black hole interiors has been the Schwarzschild geometry. From the broader perspective of the previous paragraph, however, these are a vanishingly small, non-generic and unstable class of near-singularity geometries.\footnote{A consequence of the instability is that arbitrarily small perturbations of the boundary Hamiltonian lead to $\ocal(1)$ changes of the geometry close to the black hole singularity. For example, modeling holographic CFTs in a laboratory would naturally involve noise, which destabilises the Schwarzschild singularity.} A sequence of papers over the past few years has demonstrated, in contrast, how the key ingredients for chaotic BKL dynamics can be engineered using CFT thermofield double states deformed by boundary sources or by condensates. The ingredients are Kasner epochs \cite{Frenkel:2020ysx, Hartnoll:2020rwq} and bounces between them \cite{Hartnoll:2020fhc, Cai:2020wrp, Cai:2021obq, An:2022lvo, Sword:2022oyg,DeClerck:2023fax, Caceres:2026mug}.
In the spirit of these works, our first result is that
\begin{itemize}

\item Homogeneous BKL dynamics (Misner's `mixmaster universe' \cite{Misner:1969hg}) can be realised in an AdS$_4$ M-theory black hole interior within ABJM holographic duality, by turning on sources for certain R-symmetry currents in the thermal state. This is \S\ref{sec:mtheory}.

\end{itemize}
In constructing the backgrounds we further encounter a nematic phase transition in \S\ref{sec:phase} and an explicit instance of `mass inflation' \cite{Poisson:1989zz} in \S\ref{sec:oscillation}.

The next step is to probe the BKL dynamics from the boundary. Established boundary methods to probe the interior include geodesics \cite{Fidkowski:2003nf}, quasinormal modes \cite{Festuccia:2005pi}, entanglement entropy \cite{Hartman:2013qma} and thermal one-point functions \cite{Grinberg:2020fdj}. We will deploy all of these. A further widely-studied probe of the interior, not considered here, is complexity \cite{Stanford:2014jda}. While we focus on thermofield double states, some results below will also apply to pure state black holes, as we explain in the discussion \S\ref{sec:discussion}. The interior is causally disconnected from the boundary, but remains accessible to boundary observables
whenever the bulk background is analytic (as in simple thermal states). In this vein, we will demonstrate that BKL chaos is imprinted in the subtle properties of analytically continued thermal observables. The first step in this direction is our discovery that BKL bounces have a robust consequence for analytically continued interiors:
\begin{itemize}

\item The interior BKL chaos of these backgrounds includes `magnetic' bounces where $g_{tt}$ goes from expanding to contracting. We show in \S\ref{sec:mbounce} that
\be
\text{magnetic bounce} \quad \Rightarrow \quad \text{holographic veils} \,,
\ee
where a {\it holographic veil} is a stationary point of the transverse area $\sqrt{g_{xx} g_{yy}}$, occurring at a complex value of the radial coordinate. See \S\ref{sec:veil}. 

\end{itemize}
Such veils give the title to our paper and, we will now explain, structure the boundary access to the interior in crucial ways.

Quasinormal modes (QNMs) are the fundamental excitations of dissipative systems, such as black hole exteriors \cite{Berti:2009kk}. Large overtone QNMs can be computed using a WKB analysis, which hinges on identifying contours of complexified radial coordinate along which WKB solutions are purely oscillatory. These are called anti-Stokes lines \cite{Motl:2003cd, Cardoso:2004up, Natario:2004jd}. Whenever the anti-Stokes line from the boundary to the singularity can be deformed to the real axis then the near-singularity behaviour of the physical metric is encoded in large overtone QNMs \cite{Hartnoll:2026vhu,Xiao:2026pir,Coviello:2026tol,Coviello:2026hdc}. However:
\begin{itemize}
\item In \S\ref{sec:qnmveil} we define a shear-mode perturbation of the background and show that
\be
\text{holographic veil} \quad \Rightarrow \quad \text{branch cut in the Stokes geometry} \,.
\ee
These branch cuts prevent the anti-Stokes line from being deformed to the real axis. The large overtone modes are controlled by a complex Reissner-Nordstr\"om singularity and not by the physical mixmaster dynamics. See Fig.~\ref{fig:zplane}.
\end{itemize}
Thus the physical singularity is `veiled' from the boundary QNMs, and the thermal two-point function more generally. Additionally, the veil branch cut directly impacts the QNM spectrum:
\begin{itemize}
\item In  \S\ref{sec:overdamped} we show that
\be
\text{veiled singularity} \quad \Rightarrow \quad \text{family of overdamped QNMs} \,.
\ee
The decay rate of these modes is determined by an integral around the holographic veil branch cut. See Fig.~\ref{fig:gammacut}.
\end{itemize}
These overdamped modes are in addition to the usual `christmas tree' QNMs, see Fig.~\ref{fig:qnm}.

Holographic veils are complex stationary points of the transverse area. This suggests a connection to holographic entanglement entropy \cite{Ryu:2006bv,Hubeny:2007xt}. It was explained in \cite{Hartman:2013qma} that the late-time evolution of holographic entanglement entropy is controlled by a stationary bulk surface. It is also familiar that the large-width limit of holographic entanglement is controlled by the black hole horizon \cite{Ryu:2006bv}. In \S\ref{sec:entanglement} we use the same mechanism to show that, in the $G_N \to 0$ limit:
\begin{itemize}
\item The holographic entanglement entropy $S_\text{EE}$ of a strip of width $\ell$ can be analytically continued to complex $\ell$. As $|\ell| \to \infty$ the entanglement grows linearly with $\ell$ as $S_\text{EE} \sim \frac{{\mathcal A}_\text{v}}{4 G_N} \ell $, where ${\mathcal A}_\text{v}$ is the transverse area of a holographic veil (or a horizon). In other words,
\be
\text{holographic veil} \quad \Rightarrow \quad \text{linear growth of complexified $S_\text{EE}$} \,.
\ee
\end{itemize}
The particular veil or horizon depends on the angle of the large $\ell$ limit in the complex plane and on the sheet. We call the branch cuts in the $\ell$ plane {\it entanglement folds}. We have seen that BKL bounces produce veils, hence the complexified $S_\text{EE}$ probes the BKL dynamics.

The BKL bounces can be seen even more directly by refining the one-point function probe of Grinberg and Maldacena \cite{Grinberg:2020fdj}:
\begin{itemize}
\item In \S\ref{sec:onepoint} we model the thermal one-point function $\langle \ocal \rangle$ of a high-dimension operator with a bulk scalar field $\phi$ coupled to the Weyl tensor squared, $\phi \, W_{\mu \nu \rho \sigma} W^{\mu \nu \rho \sigma}$. We analytically continue to complex scaling dimension $\Delta = \frac{3}{2} - i \mu$ and show that, as a function of $\mu$,
\be
\text{interior bounces} \quad \Rightarrow \quad \text{pairs of zeros of the one-point function} \,.
\ee
From the location of these zeros one can read off the Kasner exponents in the interior.
\end{itemize}
As we explain below, this final computation does not yield a controlled one-point function within ABJM theory. It is rather an analysis in the spirit of \cite{Myers:2016wsu}, incorporating a minimal ingredient necessary to generate thermal one-point functions. We hope that the fascinating structure that emerges in this model computation will motivate further work in this direction.

\section{Mixmaster universes in M-theory}
\label{sec:mtheory}

This section introduces a black hole background in M-theory, dual to nonzero-temperature ABJM theory with R-symmetry sources turned on. We show that the black hole interior exhibits a class of BKL dynamics identical to the classic `mixmaster' universe of Misner \cite{Misner:1969hg}.

\subsection{Consistent supergravity truncation and black hole ansatz}

In this section we recall how Einstein-Yang-Mills theory can be embedded into M-theory and we make an ansatz for magnetically charged AdS$_4$ black holes in this theory. These will be the backgrounds we use throughout. The starting point is that the following metric and four-form field strength give a consistent truncation of eleven-dimensional supergravity \cite{Mann:2006jc}, with $i = 1,2,3$ and $0 \leq \xi \leq \frac{\pi}{2}$,
\be
\begin{aligned}
ds^2_{11} & = ds^2_4 + 4 d\xi^2 + \cos^2\xi \, \sigma^i \sigma^i + \sin^2\xi \,\tau^i \tau^i \,, \\
F_4 & = - 3\, \text{vol}_4 + \left[\frac{\sin 2 \xi}{2} d\xi \wedge(\sigma^k - \tau^k) + \frac{\vep^{ijk}}{4} \left(\cos^2\xi \sigma^i \wedge \sigma^j + \sin^2\xi \tau^i \wedge \tau^j  \right)\right] \wedge \star_4 F^k \,,
\end{aligned}\label{eq:uplift}
\ee
where
\be\label{eq:omegas}
\sigma^i = \omega^i - \half A^i \,, \qquad \tau^i = \tilde \omega^i - \half A^i \,, \qquad  F^i = dA^i + \frac{\vep^{ijk}}{4} A^j \wedge A^k \,,
\ee
and the invariant $SU(2)$ forms obey $d\omega^i = - \frac{1}{2} \vep^{ijk} \omega^j \wedge \omega^k$ and $d\tilde \omega^i = - \frac{1}{2} \vep^{ijk} \tilde \omega^j \wedge \tilde \omega^k$. This is a special case of a more general truncation \cite{Cvetic:1999au}, further setting the axion and dilaton to zero and the two $SU(2)$ gauge fields to be equal.

When $A^i = 0$ in (\ref{eq:omegas}) the background (\ref{eq:uplift}) is the Freund-Rubin compactification of eleven-dimensional supergravity on $S^7$. Later, to connect to ABJM theory, we will need to consider instead $S^7/\mathbb{Z}_k$. The ansatz (\ref{eq:uplift}) describes a consistent $SU(2)$ subsector that commutes with this $\mathbb{Z}_k$ action \cite{BenettiGenolini:2018iuy}. Therefore, the quotient by
$\mathbb{Z}_k$ will play no role in our bulk computations below. The effective four-dimensional action for the $SU(2)$ subsector is \cite{Mann:2006jc}
\be\label{eq:thy}
S = \frac{1}{2 \k^2_4} \int d^4x \sqrt{-g} \left(R + 6 - \frac{1}{4} F^i_{\mu\nu} F^{i \, \mu \nu} \right) \, ,
\ee
where $\kappa_4^2 = 8 \pi G_N$.
We work in units with the AdS$_4$ radius set to one.
We will find asymptotically AdS$_4$ black hole solutions to the Einstein-Yang-Mills theory (\ref{eq:thy}) within the ansatz
\be
\begin{aligned}
ds^2_4 & = \frac{1}{z^2} \left(-s^2 f dt^2 + \frac{dz^2}{f} + e^{2 B} dx^2 + e^{-2 B} dy^2 \right) \,, \\
A^1 +i A^2 & = w_x dx + i w_y dy \,.
\end{aligned}\label{eq:ansatz}
\ee
Here $s,f,B,w_x,w_y$ are functions of $z$ only. Correspondingly, the only nonvanishing components of the field strength are
\be\label{eq:F}
F^1_{zx} = w_x' \,, \qquad F^2_{zy} = w_y' \,, \qquad F^3_{xy} = \half w_x w_y \,.
\ee
That is, we will be looking at magnetically charged black holes. Magnetically charged solutions were considered in \cite{Mann:2006jc, Grandi:2021bsp}. Closely related electrically charged solutions were constructed as duals to $p$-wave superconductors in \cite{Arias:2012py}. The interiors of the $p$-wave geometries were explored in \cite{Cai:2021obq, Sword:2022oyg}, and exhibit similar cosmological billiards to that described in \S\ref{sec:billiards}. Magnetically charged backgrounds are less prone to nonperturbative brane nucleation instabilities that are common in string-theoretic electrically charged backgrounds, e.g.~\cite{Yamada:2006rx, Hartnoll:2009ns, Henriksson:2019ifu, Choi:2024xnv}.

The equations of motion for the ansatz (\ref{eq:ansatz}) are given in Appendix \ref{ap:equations}. We will numerically construct solutions that are asymptotically AdS and have a regular non-extremal horizon. Near the AdS boundary $z \to 0$ and
\be\label{eq:asymptotic}
f \to 1 \,, \quad s \to 1 \,, \quad B \to 0 \,, \quad w_x \to W_x \,, \quad w_y \to W_y \,.
\ee
Thus there are explicit sources $W_x$ and $W_y$ for the non-abelian vector potential. The horizon will be at $z = z_{\mathcal H}$. At $z > z_{\mathcal H}$ the $z$ direction is timelike and
the approach to the interior cosmological singularity is at $z \to \infty$, as we describe in the following section. The temperature is given by
\be
T = - \frac{s(z_{\mathcal H}) f'(z_{\mathcal H})}{4 \pi} \,.
\ee

\subsection{Phase diagram}
\label{sec:phase}

In Appendix \ref{ap:equations} we describe how to construct numerical backgrounds that are regular on the horizon and have the near-boundary behaviour (\ref{eq:asymptotic}). Given the sources $W_x$ and $W_y$ for the non-abelian currents, the basic information to extract from the geometries are the expectation values for these currents and for the energy-momentum tensor. These are encoded in the leading normalisable behaviour of the variables as $z \to 0$:
\be\label{eq:nbdy}
w_a = W_a + j_a z + \ocal(z^2) \,, \quad f = 1 - f_3 z^3 + \ocal(z^4) \,, \quad B = -B_3 z^3 + \ocal(z^4) \,, \quad s = 1 + \ocal(z^4) \,.
\ee
Here $a = x,y$. The current and energy-momentum tensor expectation values are then, from the basic holographic dictionary (e.g. \cite{deHaro:2000vlm, Hartnoll:2018xxg}),
\be\label{eq:vevs}
\langle J^1_x \rangle = \frac{j_x}{2\k^2_4} \,, \quad 
\langle J^2_y \rangle = \frac{j_y}{2\k^2_4} \,, \quad 
\langle T_{tt}\rangle = \frac{f_3}{\k^2_4} \,, \quad \langle T_{xx}\rangle = \frac{f_3 - 6 B_3}{2\k^2_4} \,, \quad \langle T_{yy}\rangle = \frac{f_3 + 6 B_3}{2\k^2_4} \,.
\ee
Note that $-\langle T_{tt}\rangle+\langle T_{xx}\rangle+\langle T_{yy}\rangle = 0$, as required.

\begin{figure}[h]
    \centering
    \includegraphics[width=0.8\linewidth]{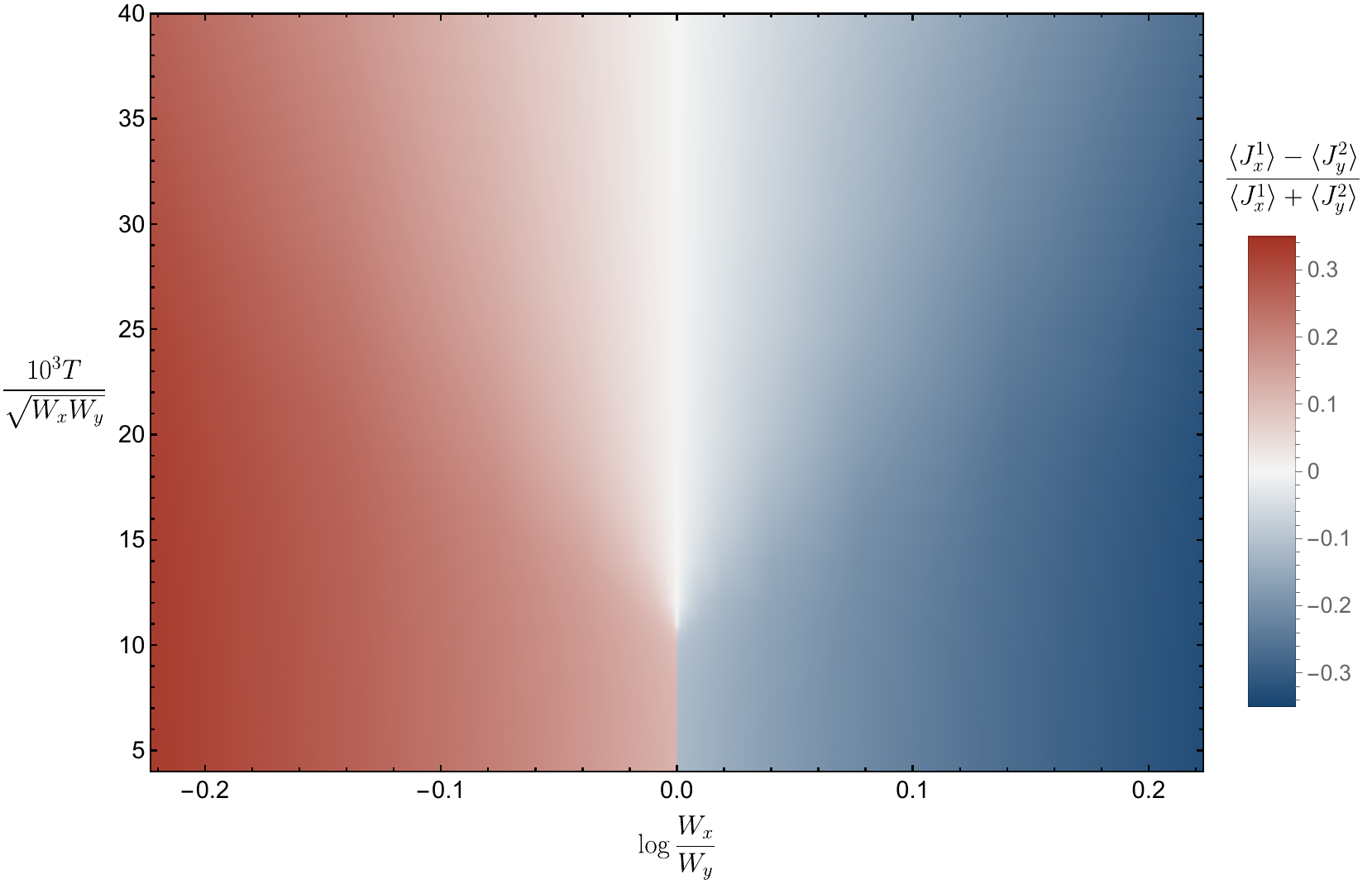}
    \caption{Phase diagram of the sourced Einstein-Yang-Mills geometries. The density plot shows the current imbalance $\frac{\langle J^1_x \rangle-\langle J^2_y \rangle}{\langle J^1_x \rangle+\langle J^2_y \rangle}$ as a function of the dimensionless temperature and the dimensionless anisotropy of sources. The logarithm on the $x$ axis ensures a reflection symmetry under $W_x \leftrightarrow W_y$. Note that the currents are negative for positive sources.}
    \label{fig:phase}
\end{figure}

Fig.~\ref{fig:phase} shows the phase diagram of this system obtained from the numerical backgrounds. There are two dimensionless tuning parameters: the temperature in units of the sources and the anisotropy in the sources. The density plot shows the resulting imbalance in the currents.
The most notable aspect of the phase diagram in the figure
arises when the sources are equal, $W \equiv W_x = W_y$.
Below the temperature $T_\text{c} \approx 0.0107 \, W$ the current imbalance is nonzero, indicating spontaneous symmetry breaking. This fact has been described previously \cite{Grandi:2021bsp} and corresponds to the onset of nematic ordering for $T < T_\text{c}$, as we now review. The background sources are $A^i_a = W \delta^i_a$, and hence preserve an $SO(2)$ symmetry that acts simultaneously on the colour and spatial indices. The current imbalance leads to a nonzero traceless, symmetric (i.e.~nematic) order parameter that breaks this symmetry
\be
\langle J^i_a \rangle \delta^i_{b} + \langle J^i_b \rangle \delta^i_{a} - \langle J^k_k \rangle \delta_{ab} = \left( \langle J^1_x \rangle - \langle J^2_y \rangle \right) 
\begin{psmallmatrix}
  1 & 0 \\
  0 & -1
\end{psmallmatrix}_{ab}
\,.
\ee
Of course, when the two sources are distinct the symmetry is broken explicitly.

We recall in Appendix \ref{ap:equations} that the underlying cause of the symmetry breaking here is understood. Namely, at $T=0$ the symmetric near-horizon  geometry develops Lifshitz scale invariance \cite{Devecioglu:2014iia} and there is a mode in this geometry with a complex scaling exponent \cite{Grandi:2021bsp}. This is a generic criterion for instability \cite{Hartnoll:2018xxg}. Spontaneous symmetry breaking in the boundary field theory is not, per se, our primary interest here. That said, the BKL dynamics we describe below only occurs in geometries where the symmetry is broken either explicitly or spontaneously. 
In \S\ref{sec:oscillation} we will see that the late-interior dynamics of symmetric backgrounds, those with $W_x = W_y$ and $T>T_\text{c}$, are instead characterised by what has been called `oscillatory mass inflation'.

\subsection{Cosmological billiards}
\label{sec:billiards}

\begin{figure}[h]
    \centering
    \includegraphics[width=0.92\linewidth]{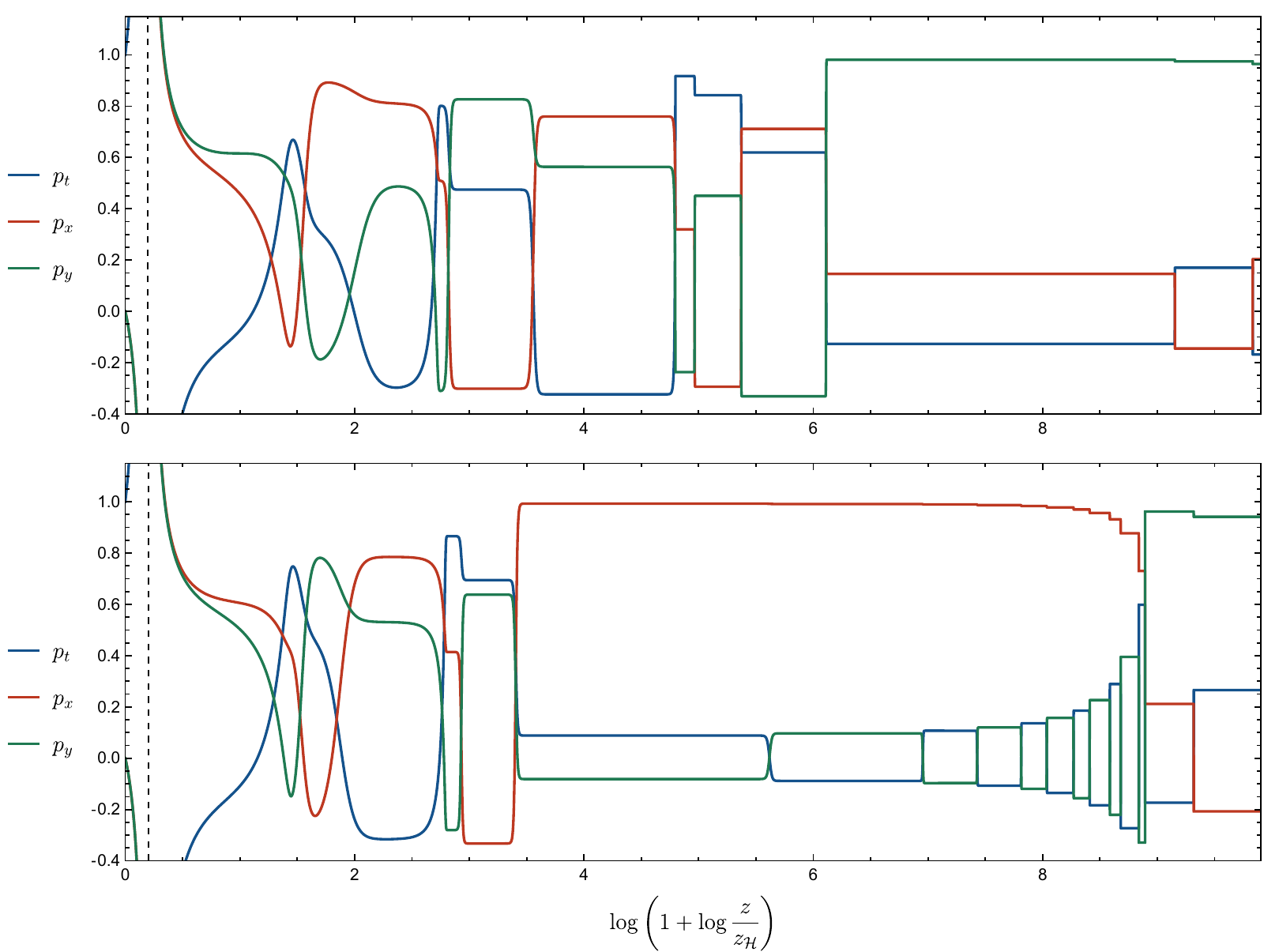}
    \caption{Interior evolution of the expansion variables (\ref{eq:pp}). The two different runs correspond to $W_x/W_y \approx 1.369$ and $T/\sqrt{W_x W_y} \approx 0.103$ (top) and to $W_x/W_y \approx 0.699$ and $T/\sqrt{W_x W_y} \approx 0.121$ (bottom). The plots start at the horizon. The dashed lines show the time of maximal spatial volume at which $q_t + q_x + q_y=0$ and hence the $p_a$ diverge. Cosmological billiard dynamics sets in once the spatial volume starts to collapse.}
    \label{fig:interior}
\end{figure}

The backgrounds constructed in the previous section can be extended to the black hole interior, with $z > z_\mathcal{H}$. To characterise the behaviour it is convenient to plot the instantaneous expansion in the spatial $a = \{t,x,y\}$ directions. This is defined as
\be\label{eq:pp}
p_a \equiv \frac{q_a}{q_t + q_x + q_y} \,, \qquad q_a \equiv \frac{d \log g_{aa}}{d \log \frac{z}{z_\mathcal{H}}} \,. 
\ee
There is no sum over repeated indices and the metric components are to be read off from (\ref{eq:ansatz}). Fig.~\ref{fig:interior}
shows the evolution of these expansions to very late interior times.
The solutions are seen to exhibit classic cosmological billiard behaviour \cite{Belinsky:1970ew, DeClerck:2023fax}. There is a sequence of so-called Kasner epochs over which the Kasner parameters (\ref{eq:pp}) are constant, punctuated by transitions between them that become sharper at later times. The lower plot in Fig.~\ref{fig:interior} further shows a so-called Kasner era, given by repeated exchanges between a pair of exponents. Over a Kasner epoch the metric (\ref{eq:ansatz}) has
\be\label{eq:kasnerform}
f = - f_o z^{3 + \b^2} \,, \qquad s = s_o z^{-\beta^2} \,, \qquad B = B_o + \beta \log z \,,
\ee
where $\{f_o,s_o\} > 0$, $B_o$ and $\beta$ are constants. The Kasner exponents are
\be
p_t=\frac{\beta^2-1}{\beta^2+3} \,,
\qquad
p_x=\frac{2(1-\beta)}{\beta^2+3} \,,
\qquad
p_y=\frac{2(1+\beta)}{\beta^2+3} \,.
\ee
These obey the familiar relations for pure gravity, $\sum_a p_a = 1$ and $\sum_a p_a^2 = 1$, because the Yang-Mills fields $\{w_x, w_y\}$ are frozen over a Kasner epoch (see also Appendix \ref{ap:hamiltonian}).

The emergence of mixmaster behaviour as $z \to \infty$ can be diagnosed without solving the equations of motion. This is most transparent in a Hamiltonian description \cite{Damour:2002et}. It is instructive to obtain the generator of evolution in $z$ within the ansatz (\ref{eq:ansatz}), done in Appendix \ref{ap:hamiltonian}. The analysis is complicated by the presence of gauge constraints. The Hamiltonian on the reduced (gauge-fixed) phase space is found to be
\be\label{eq:Hred}
H_\text{red} = \frac{48(f-1) + z^4 w_x^2 w_y^2}{16 z} \,\hat \pi_f - \frac{z^2 \hat \pi_B^2 + 4 \left(e^{2 B}\pi_x^2 + e^{-2 B} \pi_y^2 \right)}{4 z^3 f} \frac{1}{\hat \pi_f}\,.
\ee
Here $\{\hat \pi_f, \hat \pi_B, \pi_x, \pi_y\}$ are the momenta conjugate to $\{f,B,w_x,w_y\}$. The hats on some of the momenta here arise as part of the computations in Appendix \ref{ap:hamiltonian}, they are purely notational. The equations of motion for $\{f,B,w_x,w_y\}$ are obtained from (\ref{eq:Hred}) via Hamilton's equations. The variable $s$ has been eliminated during the analysis in Appendix \ref{ap:hamiltonian} and is given by $s = \frac{1}{2} z^3 \hat \pi_f$. Note that $f<0$ in the black hole interior.

To show the emergence of cosmological billiards, we perform a time-dependent canonical transformation on (\ref{eq:Hred}): $(f, B, \hat \pi_f,\hat \pi_B,z) \mapsto (u,v,\pi_u,\pi_v,\tau)$. The details of this transformation are also in Appendix \ref{ap:hamiltonian}, which also describes a trivial rescaling $(w_x,w_y,\pi_x,\pi_y) \mapsto (\hat w_x,\hat w_y,\hat \pi_x,\hat \pi_y)$. The upshot is a Hamiltonian $H_\tau$, generating time evolution in $\tau$, that describes a particle moving in the hyperbolic upper-half plane in the presence of a potential,
\be\label{eq:Htau}
H_\tau^2 = v^2 \left(\pi_u^2 + \pi_v^2 \right) + V_\tau(u,v) \,,
\ee
where $v \geq 0$, and the time-dependent potential
\be\label{eq:V}
V_\tau(u,v) \equiv e^{2 \tau} \left[\hat \pi_x^2 e^{-2(1-u) \frac{e^\tau}{2v}}
+ \hat \pi_y^2 e^{-2(1+u) \frac{e^\tau}{2v}}
+ \tfrac{1}{4} \hat w_x^2 \hat w_y^2 e^{-(u^2+v^2-1) \frac{e^\tau}{2v}} - 12 \, e^{-(u^2 + v^2 + 3) \frac{e^\tau}{2v}}\right] \,.
\ee
Towards the singularity $\tau \to \infty$. The final term in (\ref{eq:V}), which is the remnant of the cosmological constant, is always exponentially small and hence irrelevant. The remaining terms tend towards sharp `walls', i.e.~infinite step function potentials with a positive prefactor. These walls confine the hyperbolic particle dynamics in (\ref{eq:Htau}) to the domain
\be\label{eq:domain}
- 1 < u < 1 \qquad \text{and} \qquad u^2+v^2 > 1 \,.
\ee
This is a famous domain in hyperbolic space. It is an equilateral triangle on the Poincar\'e disc, reaching the boundary at three cusps, and equal to half of a fundamental domain of the modular group $\Gamma(2)$. It is also precisely the billiard domain of Misner's mixmaster cosmology \cite{Misner:1969hg, Misner:1969ae, Belinski:2017fas}.

\begin{figure}[h]
    \centering
    \includegraphics[width=0.85\linewidth]{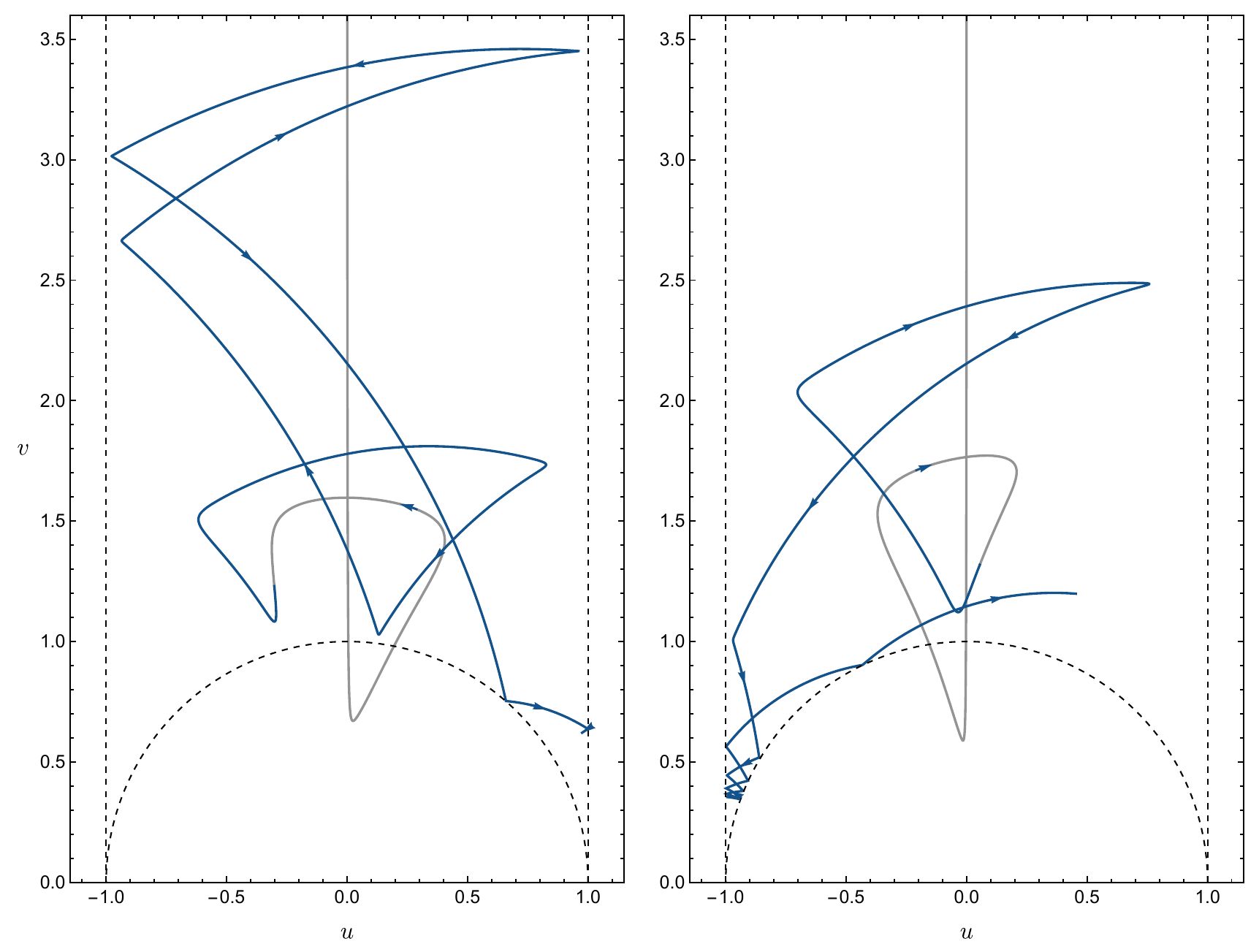}
    \caption{Billiard trajectories corresponding to the black hole interiors of Fig.~\ref{fig:interior}. The dashed lines are the domain boundaries (\ref{eq:domain}). The trajectory is plotted in grey for $\log\frac{z}{z_\mathcal{H}} < 10$ and in blue at later interior times. At late times the bounces off the domain boundaries are sharper as the billiard limit becomes more accurate. The `magnetic' bounces are off the bottom semicircle while the `electric' bounces are off the vertical right and left walls.}
    \label{fig:mixmaster}
\end{figure}

In Fig.~\ref{fig:mixmaster} we show the interiors of Fig.~\ref{fig:interior} as hyperbolic billiard trajectories on the domain (\ref{eq:domain}).
At early times the trajectory can leave the billiard domain, but at late times there are increasingly sharp bounces off the domain boundaries. The initial conditions in this plot, descending vertically at $u=0$, are set by the presence of a horizon together with the precise canonical transformation described in Appendix \ref{ap:hamiltonian}.

It is known that the full cosmological billiard of eleven-dimensional supergravity, with all components of the metric and four-form flux turned on and with generic spatial inhomogeneities, is an octonionic modular domain in $\mathbb{H}_9$\cite{Damour:2000hv, Damour:2002et,Feingold:2008ih}. The uplift in (\ref{eq:uplift}) is highly symmetric from an eleven-dimensional perspective, and hence the coefficients of some of the dominant eleven-dimensional walls vanish. Indeed, the walls obtained in (\ref{eq:domain}) descend from sub-dominant walls in eleven-dimensional supergravity.

Standard arguments, see e.g.~\cite{Damour:2002et} and Appendix \ref{ap:hamiltonian}, show that the gauge field variables $\{\hat w_x,\hat w_y,\hat \pi_x,\hat \pi_y\}$ freeze to constant values at late interior times. With the exception of fine-tuned trajectories where some of these variables tend towards zero the universe is confined within the domain (\ref{eq:domain}). Another exceptional case are trajectories that tend precisely into one of the cusps, that we turn to in the following \S\ref{sec:oscillation}.

\subsection{Oscillatory mass inflation}
\label{sec:oscillation}

The interior dynamics is different in symmetric cases where, throughout the geometry, $B=0$ and $w_x = w_y$. From the canonical transformation in Appendix \ref{ap:hamiltonian}, the fact that $B=0$ translates into the restriction that $u=\pi_u=0$. The hyperbolic particle is therefore confined to a trajectory that moves vertically up and down in the domain shown in Fig.~\ref{fig:mixmaster}. Dropping the exponentially subleading final term in the potential (\ref{eq:V}) and setting $\hat w \equiv \hat w_x = \hat w_y$ and $\hat \pi \equiv \pi_x = \pi_y$, the Hamiltonian (\ref{eq:Htau}) becomes
\be\label{eq:Hsym}
H_\tau^2 \approx v^2 \pi_v^2 + e^{2 \tau} \left[2 \hat \pi^2 e^{- \frac{e^\tau}{v}} + \tfrac{1}{4} \hat w^4 e^{-(v^2-1) \frac{e^\tau}{2v}}\right] \,.
\ee
The motion resulting from (\ref{eq:Hsym}) at late interior times has been fully characterised in \cite{Donets:1996ja, Breitenlohner:1997hm, Galtsov:1997lak}. One finds that, similarly to the mixmaster dynamics above, the gauge field and its momenta freeze at late interior times. That is, $\hat \pi$ and $\hat w$ become constant in (\ref{eq:Hsym}).
The two terms in the potential in (\ref{eq:Hsym}) can then be seen to cause oscillatory motion: the second term bounces the particle up away from $v=1$ while the first term causes the particle to turn around before reaching $v \to \infty$. It is instructive to briefly review the physics at work here.

\begin{figure}[h]
    \centering
    \includegraphics[width=0.95\linewidth]{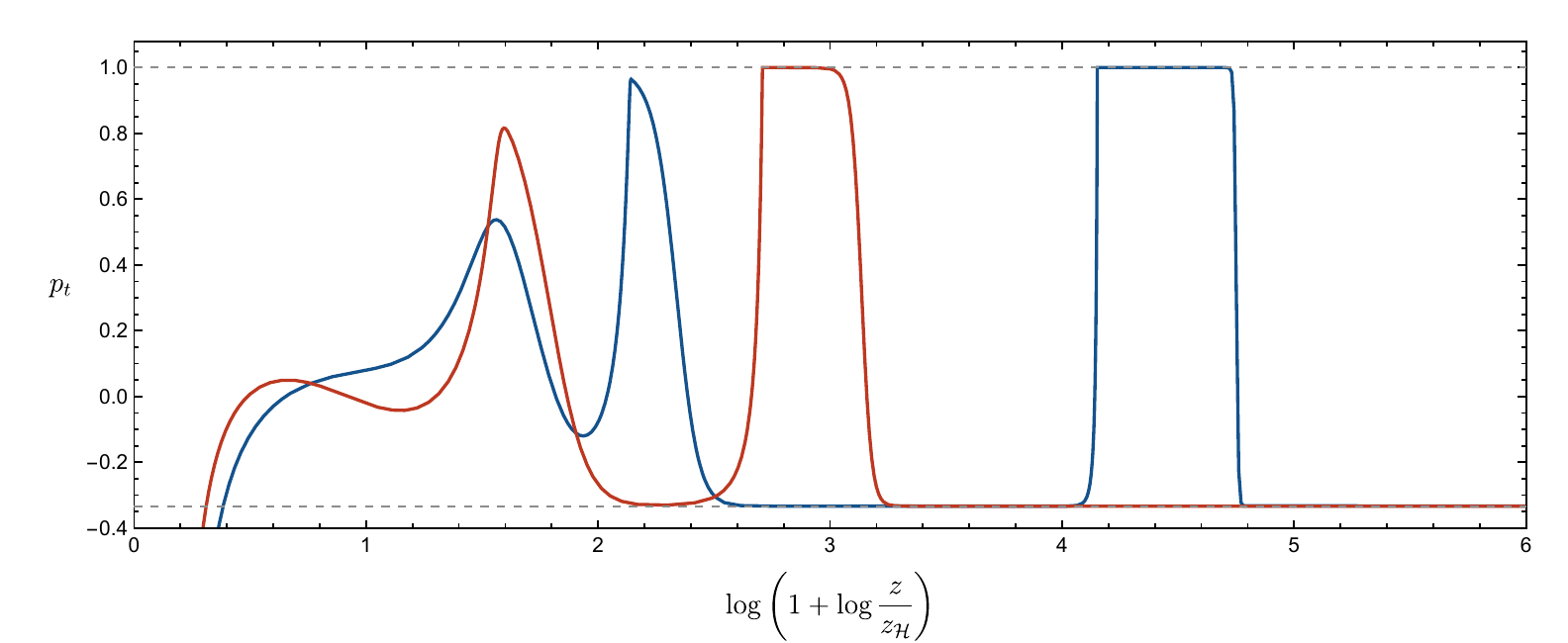}
    \caption{Two interior geometries preserving the $SO(2)$ symmetry. The red plot has $T/W \approx 0.0272$ and the blue plot has $T/W \approx 0.0523$. The dashed lines show the Schwarzschild value of $p_t = -\frac{1}{3}$ and the Reissner-Nordstr\"om inner horizon value of $p_t=1$.}
    \label{fig:sym}
\end{figure}

Two illustrative symmetric interiors are shown in Fig.~\ref{fig:sym}.
The rotational symmetry implies that $p_x = p_y = \frac{1}{2}(1 - p_t)$ and hence only $p_t$ is independent.
The far interior shows sudden oscillations between two fixed exponents, $p_t = -\frac{1}{3}$ and $p_t = 1$. Indeed, these are the only exponents consistent with the constraint $0 = p_t^2 + p_x^2 + p_y^2 -1 = p_t^2 + 2\frac{(1-p_t)^2}{4} -1 = \frac{1}{2}(p_t-1)(3 p_t+1)$.
Recall that $p_t = -\frac{1}{3}$ corresponds to a vacuum Schwarzschild-like regime while $p_t=1$ corresponds to a Reissner-Nordstr\"om inner horizon-like regime. The instability of both of these regimes, and hence the oscillations between them, can be seen easily by specialising (\ref{eq:Hred}) to this symmetric setting and again dropping the subleading term,
\be\label{eq:Hred2a}
H_\text{red} \approx \frac{3 f \hat \pi_f}{z} + \frac{z^3 w^4}{16} \hat \pi_f - \frac{ 2 \pi^2}{z^3} \frac{1}{f \hat \pi_f}\,.
\ee
Here, again, $\pi$ and $w$ are constant in the late interior.

We can solve (\ref{eq:Hred2a}) in the limit where one of the $w^4$ or $\pi^2$ terms dominates over the other, as occurs close to the transitions. In the former limit the solution is
found to be
\be\label{eq:muo}
f \approx - \mu_o z^3 + \frac{w^4}{16} z^4 \,.
\ee
The first term in (\ref{eq:muo}) is the Schwarzschild one, with constant effective `mass' $\mu_o > 0$. The second term is due to the nonabelian magnetic field. As space collapses towards the singularity, and $z$ grows, the second term becomes more important. As the two terms become comparable, we obtain $f \sim (z - z_\mathcal{I})$, which can be seen to give a transition from the Schwarzschild $p_t=-\frac{1}{3}$ to $p_t=1$. If this behaviour were to continue, there would be an inner horizon at $z_\mathcal{I}$ (where $f$ would vanish). However, we show in Appendix \ref{ap:equations} that these backgrounds cannot have inner horizons. The final term in (\ref{eq:Hred2a}) must save the day, as we now explain.

When the $\pi^2$ term in (\ref{eq:Hred2a}) dominates it is immediately seen that $f \hat \pi_f = 2 s f/z^3$ is constant. Let $\mu \equiv - f/z^3$ be a $z$-dependent effective mass, generalising $\mu_o$ in (\ref{eq:muo}).
Thus $s \mu$ is constant. Integrating the remaining Hamilton equation gives $\log \frac{\mu(z)}{\mu(z_o)} = \frac{\pi^2}{(f \hat \pi_f)^2} \left(\frac{1}{z_o^2} - \frac{1}{z^2} \right)$. We see that $\mu$ grows exponentially with the radial distance, and hence $s$ collapses exponentially. Both of these are known phenomena. The growth of $\mu$ is called `mass inflation' \cite{Poisson:1990eh} and enables the first term in (\ref{eq:muo}) to remain ahead of the magnetic flux term, avoiding the inner horizon. This dynamics drives a transition away from $p_t=1$ and back to $p_t=-\frac{1}{3}$. The collapse of $s$ causes the collapse of $g_{tt}$, called the `collapse of the Einstein-Rosen bridge' in \cite{Hartnoll:2020rwq, Hartnoll:2020fhc}.

\subsection{Field theory setup}
\label{sec:abjm}

The asymptotically AdS$_4$ M-theory solution described above admits a dual CFT interpretation within ABJM theory with gauge group $U(N)_k \times U(N)_{-k}$ \cite{Aharony:2008ug}. The basic dictionary takes the form, see e.g. 
\cite{BenettiGenolini:2018iuy} and recall that we have set the AdS radius to one,
\be\label{eq:dictionary}
\frac{1}{2 \kappa_4^2} = \frac{\sqrt{2 k} N^{3/2}}{24 \pi} \,, \qquad \qquad
\int_{S^7/\mathbb{Z}_k} \ast_{11} F_4 = \frac{128 \pi^4}{k} \,,
\ee
where the flux through $S^7/\mathbb{Z}_k$ is due to the charge carried by $N$ M2 branes.
 For general $k$, the theory has ${\cal N}=6$ superconformal symmetry. For $k=1,2$, there is an enhanced ${\cal N}=8$ superconformal symmetry \cite{Aharony:2008gk}.
The 't Hooft limit corresponds to $N, k \to \infty$ with $\lambda = \frac{N}{k}$ held fixed.

For general $k$, the R-symmetry of ABJM theory is $SU(4)$. The bulk Yang-Mills field in (\ref{eq:thy}) is dual to the current $J$ of a diagonally-embedded $SU(2) \subset SU(2) \times SU(2) \subset SU(4)$. The generators for this $SU(2)$ can be written in terms of Pauli matrices $\sigma_k$ as $T_k = - \frac{i}{2} \text{diag}(\sigma_k,\sigma_k)$. The boundary sources from (\ref{eq:asymptotic}) and the field strength from (\ref{eq:F}) therefore define the nonzero background connections and field strength, with the conventions of (\ref{eq:omegas}),
\be\label{eq:back}
A_x = W_x T_1 \,,\qquad  A_y = W_y T_2 \,, \qquad F_{xy} = \half W_x W_y T_{3} \,.
\ee
The sources in (\ref{eq:back}) give a time- and space-independent deformation of the ABJM Hamiltonian,
\be\label{eq:hamdef}
H(W_x,W_y) = H_{\text{CFT}} - W_x \int d^2x J^1_{x}(\vec x) - W_y \int d^2x J^2_{y}(\vec x) + \cdots \ \,. 
\ee
The current $J$ here is that of the undeformed CFT. The ellipses denote terms that are higher order in sources, coming from the covariant coupling of a background gauge field to the Lagrangian. For example, they will include $W^2$ terms responsible for seagull diagrams \cite{Closset:2012vp}.

When the mixmaster black hole saddle dominates the thermal ensemble, the geometry computes correlation functions in the deformed CFT. For example, for operators $\ocal_1$ and $\ocal_2$,
\be
\label{eq:correlator}
\big\langle {\cal O}_1(x_1) {\cal O}_2(x_2) \big\rangle = \frac{{\rm Tr}[e^{- H(W_x,W_y)/T} {\cal O}_1(x_1) {\cal O}_2(x_2) ]}{{\rm Tr}\,e^{- H(W_x,W_y)/T}} \,, 
\ee
where the operators are evolved using the deformed Hamiltonian (\ref{eq:hamdef}).
As usual, formulae such as (\ref{eq:correlator}) require a finite volume regulator.
As always, thermal correlations are KMS invariant. The sources lead to two dimensionless parameters $\frac{W_x}{T}$ and $\frac{W_y}{T}$. Thermal one-point functions will generically depend on these,
\be\label{eq:onep}
\langle {\cal O} \rangle = T^\Delta \, b_{{\cal O}}\left(\tfrac{W_x}{T}, \tfrac{W_y}{T}\right) \,.
\ee
Applying the OPE to the product of operators in (\ref{eq:correlator}) expresses the two-point function as a sum over one-point functions (\ref{eq:onep}), see e.g. \cite{El-Showk:2011yvt,Iliesiu:2018fao}. In the absence of sources, the convergence of the OPE sum implies the absence of certain singularities in the two-point function \cite{Komargodski:2026oob,Almheiri:2026cpv}. In the presence of sources this argument does not obviously apply. We return to this point in \S\ref{sec:timesing}.

\section{Quasinormal modes and holographic veils}
\label{sec:qnmveil}

This section is concerned with the large overtone limit of QNMs and the relation of these modes to the interior dynamics. We will focus on a particularly simple shear-mode perturbation of the background at zero spatial momentum, but the essential large overtone properties are captured by the $h(z)$ function defined in (\ref{eq:H}). This function determines optical distances in the background and will control all probes that are related to null geodesics in the high overtone limit. In particular, the holographic veils that we encounter shortly are defined intrinsically to the background geometry.

\subsection{A simple perturbation}
\label{sec:pert}

The background (\ref{eq:ansatz}) is invariant under two separate reflections, when these are combined with a gauge transformation. Firstly, $x \mapsto -x$ and $(A^1,A^2,A^3) \mapsto (-A^1,A^2,-A^3)$, and secondly, $y \mapsto -y$ and $(A^1,A^2,A^3) \mapsto (A^1,-A^2,-A^3)$. Perturbations with distinct parity under these maps cannot mix. In particular, the following
time-dependent but spatially uniform shear mode is a consistent perturbation,
\be\label{eq:pert}
\delta g_{ty} = \frac{e^{-2 B}}{z^2} v(z) e^{- i \omega t} \,, \qquad \delta A_t^2 = u_t(z) e^{- i \omega t} \,, \qquad \delta A^3_x =  u_x(z) e^{- i \omega t} \,.
\ee
The modes $\delta g_{zy}$ and $\delta A_z^2$ are also allowed, and impose constraints on this sector, but can be gauged to zero. The easiest way to get the equations for the perturbations is to obtain the quadratic action for $\{\delta g_{ty}, \delta g_{zy},\delta A_t^2,\delta A^3_x, \delta A_z^2\}$ and then vary. The system can be reduced to two first-order equations and one second-order equation. There is a single dynamical degree of freedom in this sector. To capture this fact, introduce the electric field
\be\label{eq:bigE}
\mathcal{E} \equiv \frac{u_t' - w_y' v}{s} \,.
\ee
This is found to obey the equation
\be\label{eq:Eeq}
\left(\frac{s f e^{2 B}}{w_x^2} \mathcal{E}' \right)'
+ \left(\frac{e^{2B}}{s f w_x^2} \, \omega^2
- \frac{s}{4}(1 + z^2 e^{2B} w_y^2)\right)\mathcal{E} = 0 \,.
\ee
Given $\mathcal{E}$, the components of the perturbation (\ref{eq:pert}) can be recovered from the definition (\ref{eq:bigE}) along with the further constraints
\be\label{eq:cfin}
v' = - z^2 s w_y e^{2 B} \mathcal{E} \,, \qquad u_t - w_y v + \frac{2 i \omega u_x}{w_x} = \frac{4 s f e^{2 B}}{w_x^2} \mathcal{E}' \,.
\ee

The equations (\ref{eq:bigE}), (\ref{eq:Eeq}) and (\ref{eq:cfin}) lead to four constants of integration, as expected. These can be taken to be the boundary values $\{u_t^{(0)}, u_x^{(0)}, v^{(0)}\}$ and the near-boundary coefficients in $\mathcal{E} = \mathcal{E}^{(0)} + \mathcal{E}^{(1)} z + \cdots$, with the constraint $u_t^{(0)} - W_y v^{(0)}+ 2 i \omega u_x^{(0)}/W_x = (4/W_x^2) \, \mathcal{E}^{(1)}$ from (\ref{eq:cfin}). By choosing these constants appropriately, one can compute different boundary retarded Green's functions. For example, to compute the charge density response set $u_x^{(0)} = v^{(0)} = 0$, and the constraint fixes $\mathcal{E}^{(1)} = (W_x^2/4)\, u_t^{(0)}$. Imposing infalling boundary conditions at the horizon \cite{Son:2002sd}
then fixes the remaining constant of integration $\mathcal{E}^{(0)}$. The charge density response is obtained from the on-shell action, as usual, by $\langle J^{2 t}\rangle = \delta S/\delta u_t^{(0)} = - 1/(2 \k^2_4) \, \mathcal{E}^{(0)}$. Finally, the retarded Green's function is defined via $\langle J^{2 t}\rangle \equiv G_{22}^{tt}u_t^{(0)}$, and hence
\be\label{eq:Gtt}
G_{22}^{tt}(\omega) = -\frac{1}{2 \k^2_4} \frac{W_x^2}{4} \frac{\mathcal{E}^{(0)}}{\mathcal{E}^{(1)}} \,.
\ee
In (\ref{eq:Gtt}) the role of the normalisable and non-normalisable coefficients, $\mathcal{E}^{(1)}$ and $\mathcal{E}^{(0)}$ respectively, may seem reversed relative to the usual dictionary \cite{Son:2002sd}. This phenomenon is familiar from e.g.~\cite{Herzog:2007ij} and occurs because the electric field $\mathcal{E}$ is a canonical momentum. The underlying bulk Yang-Mills field has the standard boundary conditions but the `master field' $\mathcal{E}$ acquires a so-called alternate quantisation. This means that to find the quasinormal frequencies, which are the poles of $G_{22}^{tt}(\omega)$, one must impose the Neumann boundary condition $\mathcal{E}^{(1)} = 0$.

The equation (\ref{eq:Eeq}) can be cast in a Schr\"odinger form by setting $\Psi \equiv \frac{e^B}{w_x} \mathcal{E}$. One finds
\be\label{eq:schro}
h \frac{d}{dz} \left(h \frac{d \Psi}{dz}\right) + \left(\omega^2 - V \right) \Psi = 0 \,.
\ee
Here
\begin{align}
h & \equiv s f = \sqrt{\frac{- g_{tt}}{g_{zz}}} \,, \label{eq:H} \\
V & \equiv \frac{s h}{4} \left(e^{-2 B} w_x^2 + z^2 w_x^2 w_y^2 \right) + \left[h \frac{d}{dz}\left(h \frac{d}{dz} \frac{e^B}{w_x} \right)\right] \frac{w_x}{e^B} \,. \label{eq:VV}
\end{align}
Eq.~(\ref{eq:schro}) will be the basis of the remainder of this section.
The Neumann boundary condition for $\mathcal{E}$ translates, using the asymptotics (\ref{eq:nbdy}), into the mixed boundary condition at $z=0$:
\be\label{eq:mixed}
W_x \Psi'(0) + j_x \Psi(0) = 0 \,.
\ee

The relation of $h$ to metric components in (\ref{eq:H}) implies that the Schr\"odinger coordinate
\be\label{eq:X}
X(z) \equiv \int_0^z \frac{dz}{h} \,, 
\ee
is the `optical distance' of the point $z$ from the boundary. This is the boundary time taken by a radial null geodesic to reach that point. The optical distance (\ref{eq:X}) plays a fundamental role in WKB analyses of the Schr\"odinger equation (\ref{eq:schro}). The key step, it turns out, is to understand (\ref{eq:X}) in the complex $z$ plane.

\subsection{Geodesics and the complexified geometry}

As we have just explained, the quasinormal frequencies $\omega_n$ are obtained by solving (\ref{eq:schro}) with the boundary condition (\ref{eq:mixed}) and with infalling boundary conditions at the horizon. The large overtone frequencies, which have $|\omega_n| \to \infty$, can be obtained by a WKB analysis building on the original papers \cite{Motl:2003cd, Cardoso:2004up, Natario:2004jd}. We will follow the exposition in \cite{Hartnoll:2026vhu}. The infalling wave near the horizon is
\be\label{eq:hor}
\Psi_\text{hor}(z) = e^{i \omega X(z)} \,.
\ee
The solution that obeys (\ref{eq:mixed}) near the boundary is instead, noting that $X(z) \approx z$ there,
\be\label{eq:bdy}
\Psi_\text{bdy}(z) = j_x \sin\Big(\w X(z)\Big) - \omega W_x \cos\Big(\w X(z)\Big) \,.
\ee
Following the usual WKB logic, a quantisation condition giving the QNMs $\omega = \omega_n$ is obtained by matching the near-horizon solution (\ref{eq:hor}) to the near-boundary solution (\ref{eq:bdy}). However, when $\omega$ is complex this matching must be done along anti-Stokes lines $\Gamma$, which obey
\be\label{eq:as}
\text{Im} \, \Big( \omega X_\Gamma(z) \Big) = 0 \,.
\ee
Thus the $\Gamma$ are contours in the complex $z$ plane along which the WKB solution is oscillatory. The notation $X_\Gamma(z)$ means that the integral in (\ref{eq:X}) is taken along the contour $\Gamma$.

For Schwarzschild-AdS and for the black holes considered in \cite{Hartnoll:2026vhu}, which were deformed by boundary sources, the anti-Stokes lines connecting the boundary to the horizon go via the singularity at $z=\infty$. That is $\Gamma = \Gamma_{1} \cup \Gamma_{2}$, where
\be\label{eq:ggamma}
\begin{aligned}
\Gamma_{1}: \;\; & \text{bdy} \;\; \rightarrow \;\; \text{sing} \\\Gamma_{2}: \;\; & \text{sing} \;\;  \rightarrow \;\; \text{hor} \,.
\end{aligned}
\ee
The near-horizon solution (\ref{eq:hor}) continues oscillating with only the positive frequency along the anti-Stokes line $\Gamma_2$. The WKB expansion breaks down at the singularity, due to the singular behaviour of the potential $V$ in the Schr\"odinger equation (\ref{eq:schro}). This fact leads to a monodromy in connecting the first and second parts of the contour in (\ref{eq:ggamma}). The monodromy is just a generalisation of the textbook WKB matching procedure. The quantisation condition is determined by this monodromy together with the usual WKB phase factor arising from the $\Gamma_1$ part of the contour between the boundary and the singularity
\be\label{eq:Z}
{\mathcal Z}_\text{QNM} \equiv X_{\Gamma_1}(\infty) = \int_{\Gamma_1} \frac{dz}{h} \,.
\ee
The asymptotic spacing of the QNMs is then found to be
\be\label{eq:asymp}
\omega_n = \frac{\pi n}{\mathcal{Z}_\text{QNM}} + \cdots \,.
\ee
Here $\mathcal{Z}_\text{QNM}$ is complex, so the modes extend diagonally into the lower half frequency plane.

In the examples considered in \cite{Hartnoll:2026vhu}, the contour $\Gamma_1$ could be deformed to the real $z$ axis without passing through any singular points of $h$ in the complex $z$ plane. This allowed us to write
\be\label{eq:Zr}
{\mathcal Z}_\text{QNM} = {\mathcal Z}_{\R} \equiv X_{\R^+}(\infty) = \int_0^\infty \frac{dz}{h} \,.
\ee
The final integral is along the real axis, with a standard contour prescription for avoiding the zero of $h$ at the horizon. The expression (\ref{eq:Zr}) is the optical depth of a null geodesic that falls from the boundary to the singularity. This is the quantity that controls the bouncing geodesic singularity of \cite{Fidkowski:2003nf} and is the first signature of the singularity in the boundary theory. More generally, however, the deformation from (\ref{eq:Z}) to (\ref{eq:Zr}) may not be possible. In the remainder of this section we will show that mixmaster bounces, of the kind we saw in \S\ref{sec:billiards}, produce branch cuts in $h(z)$ that indeed prevent this deformation. We will call these cuts holographic veils, as they obstruct access to the physical singularity from the boundary theory.

\subsection{Holographic veils}
\label{sec:veil}

We have already noted that a simple zero of $h$ arises at the horizon. A simple zero sends the $X(z)$ coordinate in (\ref{eq:X}) to infinity. Holographic veils arise instead from branch cuts in $h$ in the complex $z$ plane. The simplest case is a square root branch cut. Suppose that close to some $z_\text{v}$ we have $h \sim \sqrt{z - z_\text{v}}$. The point $z_\text{v}$ now corresponds to some finite $X_\text{v} \equiv X(z_\text{v})$ and from (\ref{eq:X})
\be\label{eq:veil}
z - z_\text{v} = \frac{1}{2} \left. \frac{d^2z}{dX^2}\right|_{X_\text{v}} (X - X_\text{v})^2 + \cdots \,.
\ee
Thus at a holographic veil the derivative $\frac{dz}{dX} = 0$ while $X$ remains finite. Eq.~(\ref{eq:veil}) says that the $X$ plane in the vicinity of the veil is locally a double cover of the $z$ plane. We will see shortly that, in the presence of veils, the anti-Stokes line $\Gamma$ can end up on a sheet of the full complex geometry that is distinct from the sheet containing the physical singularity at $z=\infty$.

Holographic veils are defined exclusively in terms of the background geometry. They are independent of the Schr\"odinger potential and turning points of the wave equation (\ref{eq:schro}). Instead, the veils determine the multi-sheeted complex manifold on which waves propagate and into which the WKB anti-Stokes lines extend. We will now give a geometric characterisation of veils as extremal surfaces. We will use this fact in \S\ref{sec:entanglement} to show that holographic veils can be detected by analytic continuation of entanglement entropy.

In terms of the optical distance coordinate (\ref{eq:X}) the metric (\ref{eq:ansatz}) becomes
\be\label{eq:metricX}
d s_4^2 = {1 \over z^2} \Big( s h \big[-dt^2 + dX^2\big] + e^{2 B} dx^2 + e^{-2B} dy^2  \Big) \ .
\ee
Here $z,s,h,B$ are viewed as functions of $X$. While $h \to 0$ at a veil, it is not a horizon and hence the product $sh$ remains finite. Indeed, the equations of motion give $s h = \frac{8 z (h^2)' - 48 h^2}{z^4 w_x^2 w_y^2 - 48}$, and the right hand side is seen to remain finite when $h \sim \sqrt{z - z_\text{v}}$. In particular, the norm of the timelike Killing vector $\xi \equiv \pa_t$ is
\be
\xi^2 = g_{tt} = - \frac{sh}{z^2} \,,
\ee
and does not vanish at $X_\text{v}$. Keeping $sh$ finite while $h \to 0$ of course requires $s \to \infty$. Recall that many conventional black hole metrics, such as Schwarzschild or Reissner-Nordstr\"om, have $s=1$ and therefore cannot have veils.

The metric (\ref{eq:metricX}) furthermore has two spacelike Killing vectors $\pa_a$, with $a=x,y$. These have an associated area density
\be\label{eq:calA}
{\mathcal A} \equiv \sqrt{\det g_{ab}} = \frac{1}{z^2} \,.
\ee
At the AdS boundary ${\cal A} \to \infty$, whereas at the singularity ${\cal A} \to 0$. Recalling that $dz = h \, dX$ and that $X$ remains a regular coordinate at the veil, we see that veils are stationary points of the area density defined by the spatial Killing fields,
\be\label{eq:dA}
d {\cal A} = - \frac{2 h}{z^3} dX = 0 \,.
\ee

\subsection{Magnetic bounces lead to holographic veils}
\label{sec:mbounce}

To see that bounces lead to veils we can solve the equations of motion for a single late-interior bounce exactly \cite{Belinsky:1970ew, DeClerck:2023fax}. More precisely, consider a so-called `magnetic' bounce. This is caused by the $w_x^2 w_y^2$ wall in (\ref{eq:Hred}), which is the lower semicircle wall in Fig.~\ref{fig:mixmaster}. The other two `electric' walls are exponentially subleading during a magnetic bounce and can be dropped. At the end of this subsection we will see that electric bounces, in contrast, do {\it not} lead to veils. The cosmological constant term is also exponentially subleading at late interior times. Thus the Hamiltonian (\ref{eq:Hred}) can be simplified to
\be\label{eq:Hred2}
H_\text{mag} = \frac{3 f \hat \pi_f}{z} + \frac{z^3 w_x^2 w_y^2 \,\hat \pi_f}{16} - \frac{\hat \pi_B^2}{4 z f \hat \pi_f} \,.
\ee

It is immediate from (\ref{eq:Hred2}) that $w_x,w_y$ and $\hat \pi_B$ are constant over the bounce. The Hamilton equations for $f$ and $\hat \pi_f$ following from (\ref{eq:Hred2}) can be solved as follows. Recalling that $s = \frac{1}{2} z^3 \hat \pi_f$, set $q \equiv h/z^3$ and then the solution is found to be
\be\label{eq:bounce}
\left(\frac{z}{z_\text{m}}\right)^2 = 
\frac{4}{(q_{\rm out}-q_{\rm in})^2} 
\left[\frac{(q_{\rm out}-q)^{q_{\rm out}}}{(q-q_{\rm in})^{q_{\rm in}}}\right]^{2/(q_{\rm out}-q_{\rm in})} \,.
\ee
Here $z_\text{m}, q_\text{in}$ and $q_\text{out}$ are constants of integration, with the relation that $16 \, q_\text{in} q_\text{out} = \hat \pi_B^2$. The physical range is $q_\text{in} < q < q_\text{out}$, with $q<0$ throughout as $f<0$ in the interior. It is also useful to note that $E \equiv -2(q_\text{in} + q_\text{out})$ is the conserved quantity $E = z H_\text{mag} - 4 f \hat \pi_f$. To see that (\ref{eq:bounce}) describes a bounce located around $z_\text{m}$, note that for $z \ll z_\text{m}$ we have $q \to q_\text{in}$, while for $z \gg z_\text{m}$ we have $q \to q_\text{out}$. These constant values of $q$ correspond to the incoming and outgoing Kasner epochs: recall from (\ref{eq:kasnerform}) that over a Kasner epoch $q = sf/z^3 = - f_o s_o$.\footnote{The incoming and outgoing values of $q$ are related by the inversion $q_\text{out} = \hat \pi_B^2/(16\, q_\text{in})$. The equation of motion for $B$ following from (\ref{eq:Hred2}) can be written $z B' = - \hat \pi_B/(4 \, q)$. It follows that
the exponent $\beta$ in (\ref{eq:kasnerform}) obeys $\beta_\text{in/out} = - \hat \pi_B/(4 \, q_\text{in/out})$ and hence this exponent obeys the inversion $\beta_\text{out} = 1/\beta_\text{in}$. This is the transformation law between Kasner epochs related by a magnetic bounce. In terms of the standard Khalatnikov-Lifshitz parameter $u$, we have $q_{\text{in}}/q_{\text{out}}=\left((u+1)/(u-1) \right)^2$. The bounce geometry is discussed further in \S\ref{sec:imprint}}

To look for possible veils, consider the differential of the area density (\ref{eq:calA}),
\be\label{eq:dlogA}
d\log\mathcal A = \left[\frac{q_\text{in}}{q-q_\text{in}} -\frac{q_\text{out}}{q-q_\text{out}} \right] \frac{2\, dq}{q_\text{out} - q_\text{in}} \,.
\ee
The right hand side is strictly negative over the physical range $q_\text{in} < q < q_\text{out}$. Thus the collapse of the universe is strong enough to cause the area density to decrease monotonically, even while
one of the transverse directions starts expanding (at a magnetic bounce, $g_{tt}$ goes from expanding to collapsing while one of $g_{xx}$ or $g_{yy}$ go from contracting to expanding).

We have noted that throughout the physical range $q<0$. The stationary point of (\ref{eq:dlogA}) is instead at $q=0$. To see that this gives a veil at a complex $z_\text{v}$, we can expand (\ref{eq:bounce}) at small $q$ to obtain a quadratic term
\be
\frac{z}{z_\text{m}} = \frac{z_\text{v}}{z_\text{m}} \left(1 + \frac{q^2}{2 q_\text{in} q_\text{out}} + \cdots \right) \,.
\ee
Clearly this last equation leads to $h \sim q \sim \sqrt{z-z_\text{v}}$ and hence a veil of the form (\ref{eq:veil}). Setting $q=0$ in (\ref{eq:bounce})
gives the location of the veils to be, with $n \in \Z$,
\be\label{eq:bounceveils}
z^{(n)}_\text{v} = z_\text{m} \frac{2 \sqrt{q_\text{in} q_\text{out}}}{q_\text{out} - q_\text{in}} \left(\frac{q_\text{in}}{q_\text{out}} \right)^{-\frac{q_\text{in} + q_\text{out}}{2(q_\text{out} - q_\text{in})}} e^{- \frac{2\pi i \, q_\text{out}}{q_\text{out} - q_\text{in}} \left(n + \frac{1}{2}\right)} \,.
\ee
The easiest way to get the phase correctly here is to consider the logarithm of (\ref{eq:bounce}).
In (\ref{eq:bounceveils}) we see that, for generic irrational $\frac{q_\text{out}}{q_\text{out} - q_\text{in}}$, there is a dense circle of veils in the complex plane. However, it will be important to keep track of the sheet that the veils lie on. The original physical sheet is the one containing the real positive $z$ axis.

The holographic veils predicted by (\ref{eq:bounceveils}) assume that the bounce solution (\ref{eq:bounce}) remains a parametrically valid approximation to the exact solution when continued off the real $z$ axis. To verify that this is the case, we found holographic veils in the numerical backgrounds that were constructed in \S\ref{sec:billiards}. The formula (\ref{eq:bounceveils}) is found to give an excellent agreement with veils computed by numerically solving the equations of motion in the complex $z$ plane. This agreement is shown in Table \ref{tab:bounces} and establishes that the veils are indeed associated to the bounces.

\begin{table}[h]
 \centering\footnotesize
 \setlength{\tabcolsep}{5pt}
\begin{tabular}{cccccc@{}}
\toprule
bounce & $n$ & $\log z_{\rm m}$ & $q_{\rm in}/q_{\rm out}$
& $\log z_{\rm v}^{\rm pred}$ & $\log z_{\rm v}^{\rm num}$ \\
\midrule
M1 & $0$ & 2.34     & 54       & $3.13+0.06\,i$        & $3.210641+0.2142999\,i$ \\
M2 & $0$ & 14.51915 & 26.81432 & $15.37771+0.1216996\,i$ & $15.37771+0.1216996\,i$ \\
M3 & $0$ & 1047.503 & 1.480897 & $1050.137+6.532773\,i$  & $1050.137+6.532773\,i$  \\
M4 & $0$ & 2471.560 & 1.630756 & $2473.978+4.980675\,i$  & $2473.978+4.980675\,i$  \\
M5 & $0$ & 3902.259 & 1.913094 & $3904.403+3.440602\,i$  & $3904.403+3.440602\,i$  \\
M6 & $0$ & 5354.728 & 2.628063 & $5356.494+1.929651\,i$  & $5356.494+1.929651\,i$  \\
M7 & $0$ & 6898.584 & 6.966583 & $6899.757+0.5265313\,i$ & $6899.757+0.5265313\,i$ \\
M7 & $1$ & 6898.584 & 6.966583 & $6899.757+1.579594\,i$  & $6899.757+1.579594\,i$  \\
\bottomrule
\end{tabular}
\caption{Numerically computed holographic veils compared to the prediction (\ref{eq:bounceveils}) from the bounce solution. The real geometry is the lower plot in Fig.~\ref{fig:interior}. To use the prediction (\ref{eq:bounceveils}) the ratio $q_{\rm in}/q_{\rm out}$ was extracted from the conserved quantities (over each bounce) $\hat \pi_B$ and $E$, defined in the main text. The bounce location $z_\text{m}$ was then obtained by fitting to (\ref{eq:bounce}). This fit purely uses data on the real $z$ axis. The formula (\ref{eq:bounceveils}) is then seen to predict the location of the veils in the complex $z$ plane to many decimal places (with a larger error for the first bounce). In the bulk we use units with $z_\mathcal{H} = 1$. Numerically we also found veils not directly associated to bounces. Examples of these were at $z_\text{v} \approx -0.6590 + 1.0025 i$ and $z_\text{v} \approx -0.5565 + 0.4640 i$.
\label{tab:bounces}}
\end{table}

Finally, at an electric bounce, which corresponds to bounces off the vertical walls in Fig.~\ref{fig:mixmaster}, the Hamiltonian (\ref{eq:Hred}) becomes
\be
H_\text{elec} = \frac{3 f \hat \pi_f}{z} - \frac{z^2 \hat \pi_B^2 + 4 e^{\pm 2 B}\pi_{x/y}^2 }{4 z^3 f \hat \pi_f} \,.
\ee
This Hamiltonian only depends on the combination $f \hat \pi_f$ and hence $q \equiv h/z^3 = \frac{1}{2} f \hat \pi_f$ is constant throughout the bounce. There are no zeros of $q$, and hence no veils, associated to these bounces.

Only bounces off the lower semi-circle wall produce veils because the diagonal ansatz in (\ref{eq:ansatz}) has fixed the Kasner frame (see e.g.~\cite{Damour:2002et} for a definition). In more general BKL dynamics there will not be a one-to-one correspondence between walls and boundary directions. However, in the discussion around Fig.~\ref{fig:mixmaster} above we saw how the first bounce was always off the lower wall due to the way the cosmological billiard regime was entered through a horizon. It seems plausible that it will more generally be the case that the very first bounce is associated to the boundary time components of the metric, $g_{tt}$, and hence produces a veil.

\subsection{A complex magnetic singularity controls the large overtones}
\label{sec:magsing}

The equations of motion in Appendix \ref{ap:equations} admit the following large $|z|$ expansion
\be
\begin{aligned}
s & = s_\infty + \ocal\left(1/z^2 \right) \,, \quad  && B  = B_\infty + \frac{C_\infty}{z} + \ocal\left(1/z^2\right)\,, \\
w_a & = w_{a\infty} + \ocal\left(1/z^2 \right) \,, \quad && f =   \frac{w_{x\infty}^2w_{y\infty}^2}{16} z^4 + f_\infty z^3 +  \ocal(z^2) \,.
\end{aligned}\label{eq:magsing}
\ee
Beyond the orders shown, these expansions admit eight unfixed constants in general. Using the expansion (\ref{eq:magsing}) in the metric (\ref{eq:ansatz}) gives the near-singularity geometry of a magnetic Reissner-Nordstr\"om black hole. The magnetic field comes from $F^3_{xy}$ in (\ref{eq:F}). For this geometry to arise along the real $z$ axis there would need to be an inner horizon because, on the real axis, $f>0$ in (\ref{eq:magsing}). In Appendix \ref{ap:equations} we have proven that there can be no inner horizons within the metric ansatz (\ref{eq:ansatz}). As we have seen, the real $z$ interior is instead characterised by mixmaster chaos. However, along a complex $z$ contour it is perfectly possible to reach the asymptotic geometry given by (\ref{eq:magsing}). In general, the constants of integration, such as $w_{x\infty}$, will also be complex. This last fact furthermore explains why along a complex $z$ contour the geometry can avoid entering a mixmaster regime: the three walls in the potential (\ref{eq:V}) will not have positive definite coefficients, and hence they need not be repulsive.

\begin{figure}[h]
    \centering
    \includegraphics[width=0.65\linewidth]{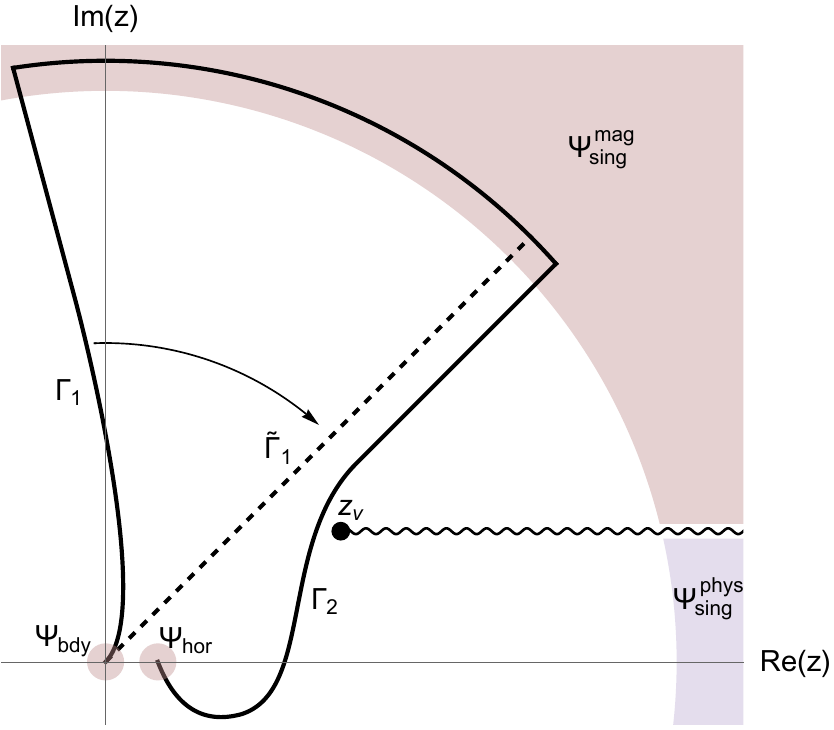}
    \caption{The anti-Stokes lines $\Gamma_1$ and $\Gamma_2$, respectively, connect the boundary and the horizon to the large $|z|$ region, where they can be matched using the near-singularity solution $\psi^\text{mag}_\text{sing}$. This singularity is a complex geometry supported by magnetic flux, described in \S\ref{sec:magsing}, and is distinct from the physical mixmaster singularity at large real $z$. The two singularities are separated by a branch cut emanating from a holographic veil at $z_\text{v}$. This branch cut prevents $\Gamma_1$ from being deformed to the real $z$ axis. Instead we deform it to a straight line $\tilde \Gamma_1$, which runs at 45 degrees.}
    \label{fig:zplane}
\end{figure}

The large $|z|$ behaviour in (\ref{eq:magsing}) can coexist with large $z$ mixmaster dynamics if there is a branch cut separating the two asymptotics. Precisely such a branch cut emerges from the first $n=0$ holographic veil in (\ref{eq:bounceveils}). This cut and the two different near-singularity asymptotics are illustrated in Fig.~\ref{fig:zplane}.
To obtain the large overtone QNMs, then, we must determine on 
which large $|z|$ sheet the anti-Stokes lines $\Gamma_1$ and $\Gamma_2$ in (\ref{eq:ggamma}) end. We have done this numerically for the geometry with bounces and veils shown in Table \ref{tab:bounces}. The result, shown schematically in Fig.~\ref{fig:zplane}, is that the contours go to the complex magnetic singularity.\footnote{To find the anti-Stokes line $\Gamma_1$ one has to know the leading-order asymptotic QNM frequencies $\omega_n$ in (\ref{eq:asymp}), which in turn depend on $\mathcal{Z}_\text{QNM}$. This may appear circular. However, $\mathcal{Z}_\text{QNM}$ can be evaluated along any contour that is homotopic to $\Gamma_1$. The strategy is to consider trial contours, and for each one obtain $\mathcal{Z}_\text{QNM}$ and hence the would-be $\omega_n$. The computation is self-consistent if the resulting anti-Stokes line is indeed homotopic to the trial contour.}

With the anti-Stokes lines shown in Fig.~\ref{fig:zplane}, the computation of the asymptotic QNMs --- including subleading corrections --- proceeds exactly as in \cite{Hartnoll:2026vhu}. The first step is to determine the potential in the
Schr\"odinger equation (\ref{eq:schro}) close to the magnetic singularity. This region determines the monodromy associated to the large $|z|$ arc in Fig.~\ref{fig:zplane}. This arc runs through an angle of $\frac{\pi}{3}$ and, in turn, controls the QNM quantisation condition. The  
Schr\"odinger equation can be put in its canonical form by introducing a coordinate $x$ such that $h\frac{d}{dz} = \frac{d}{dx}$. This coordinate can be chosen so that $x \to 0$ as $z \to \infty$. The Schr\"odinger potential (\ref{eq:VV}) is then found to have the expansion
\be\label{eq:Vexp}
V(x) = \frac{4}{9 x^2} + \frac{c_1}{x^{5/3}} + \frac{c_2}{x^{4/3}} + \frac{c_3}{x} +  \cdots  \,.
\ee
The constants $c_1, c_2, c_3$ here are fixed by the coefficients appearing in the expansion (\ref{eq:magsing}). These can be determined numerically for a given background.

Proceeding as in \cite{Cardoso:2004up, Natario:2004jd}, the leading order QNMs are
\be\label{eq:wn0}
\w_{n0} = \frac{\pi}{\mathcal{Z}_\text{QNM}} \left( n - \frac{1}{4} + \frac{\log 3}{4 \pi} i \right) \,.
\ee
These follow from the monodromy of the Bessel functions that solve the Schr\"odinger equation with the leading order potential in (\ref{eq:Vexp}), via the quantisation condition $1 - i \sqrt{3} e^{2 i \omega \mathcal{Z}_\text{QNM}} = 0$. Now proceeding as in \cite{Hartnoll:2026vhu} the leading non-analytic correction is found to be
\be\label{eq:analyt}
 \omega_n= \omega_{n0}
 +\frac{1}{2 {\mathcal{Z}}_\text{QNM}}\left[\frac{e^{-i \pi/3}\pi\Gamma(\frac{1}{6})}
 {i \sqrt3\,\Gamma(\frac{2}{3})\Gamma(\frac{5}{6})}
 \left(\frac{c_2}{3}+\frac{3 c_1^2}{8}\right)
 \frac{1}{\omega_{n0}^{2/3}}
 +\ocal\left(\frac{\log\omega_{n0}}{\omega_{n0}}\right)\right] \,.
\ee
The coefficients here are the ones appearing in the expansion of the potential (\ref{eq:Vexp}).

In Appendix \ref{app:qnm} we have checked that (\ref{eq:analyt}) is consistent with the asymptotic QNMs computed using an exterior numerical background. The numerical frequencies are shown in Fig.~\ref{fig:qnm} below. In evaluating $\mathcal{Z}_\text{QNM}$ we found it convenient to integrate along the straight line contour $\widetilde \Gamma_1$ shown in Fig.~\ref{fig:zplane}. This contour goes from the boundary to the complex magnetic singularity. In contrast, using
the optical depth ${\mathcal{Z}}_{\R}$ in (\ref{eq:Zr}), which is integrated along the real axis towards the mixmaster singularity, does not reproduce the numerically observed spacing of the asymptotic modes. The inequivalence of the two integrals $\mathcal{Z}_\text{QNM}$ and ${\mathcal{Z}}_{\R}$ is due to the branch cut emanating from the holographic veil in Fig.~\ref{fig:zplane}. The large overtone QNMs do not probe the physical BKL singularity, but rather the complex singular geometry (\ref{eq:magsing}).

\begin{figure}[h]
    \centering
    \includegraphics[width=0.55\linewidth]{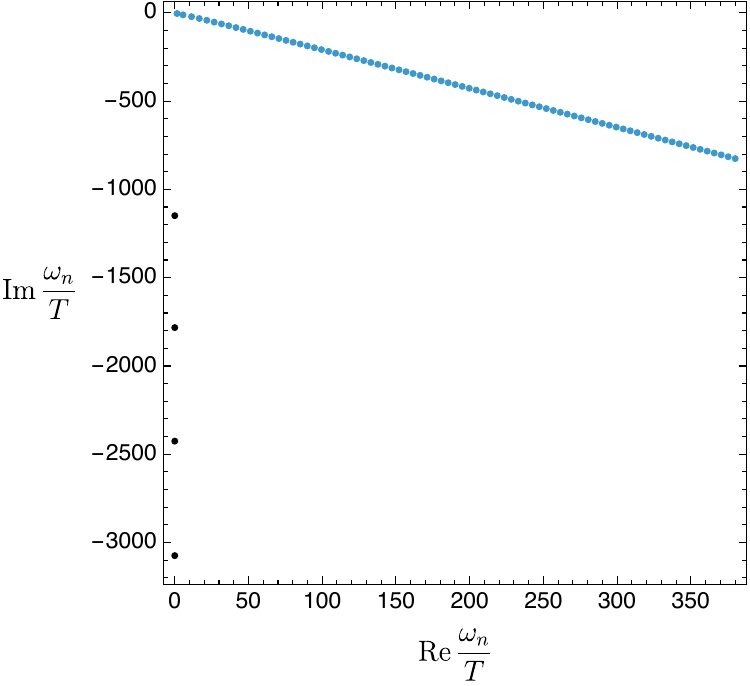}\hspace{5mm}
    \includegraphics[width=0.315\linewidth]{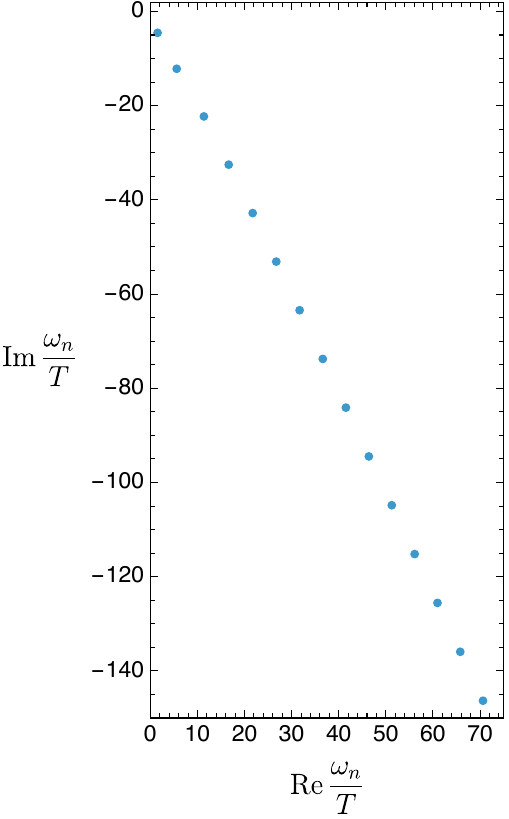}
    \caption{Numerical QNM obtained from the exterior of the lower geometry in Fig.~\ref{fig:interior}. Left plot: the first 80 `diagonal' branch modes are shown together with the first 4 overdamped modes. The real and imaginary axes have different scales on the plot, to save vertical space. Right plot: zoom in to the first 15 diagonal modes.}
    \label{fig:qnm}
\end{figure}

Large overtone QNMs controlled by complex endpoints rather than the physical singularity have recently also been described in \cite{Shi:2026cfq, Oh:2026qcm}.
Both of those geometries can be seen to have holographic veils, as we have defined them. Complex singularities also control the large overtone QNMs in `regular black holes' with no physical singularity \cite{Chen:2026efa}.

\subsection{Overdamped quasinormal modes}
\label{sec:overdamped}

We have found that in addition to the QNMs discussed in the previous \S\ref{sec:magsing} the geometry admits a family of purely imaginary QNMs with $\omega_j = - i \gamma_j$. These are also shown in Fig.~\ref{fig:qnm}. In this section we will explain that these overdamped modes are a consequence of the holographic veil branch cut. First, we can note that the asymptotic spacing and offset of the modes can be predicted using the thermal product formula \cite{Dodelson:2023vrw}. This formula expresses the boundary-to-boundary (or left-to-right) Green's function $G_{LR}(\omega)$ in terms of the QNMs. Assuming the presence of an additional overdamped family, this gives
\begin{equation}
 \frac{G_{LR}(\omega)}{G_{LR}(0)}=
 \prod_n\frac{1}{(1-\omega^2/\omega_n^2)
                       (1-\omega^2/\omega_n^{*2})}
 \prod_j\frac{1}{1+\omega^2/\gamma_j^2} \,.
 \label{eq:twoproduct}
\end{equation}
We have already seen that asymptotically $\omega_n \approx \omega_{n0}$. Now we further posit that, with $j\geq1$,
\be\label{eq:gamma}
 \gamma_j \approx \frac\pi{{\mathcal Z}_{\text{I}}}(j+j_0) \,.
\ee
Plugging these forms into (\ref{eq:twoproduct}) and taking the large $\omega$ limit gives
\be
 \begin{aligned}
 \prod_{n=0}^\infty
 \frac{1}{(1-\omega^2/\omega_n^2)(1-\omega^2/\omega_n^{*2})}
 &\propto \omega^{-3}e^{-2\operatorname{Im}{\mathcal Z}_{\text{QNM}}\,\omega},\qquad
 \prod_{j=1}^\infty\frac{1}{1+\omega^2/\gamma_j^2}
 \propto \omega^{1+2 j_0}e^{-{\mathcal Z}_{\text{I}}\omega}.
 \end{aligned} \label{eq:tomatch}
\ee
These limits can be found, for example, by writing the product as the exponential of a sum and approximating by an integral. Alternatively, by using the product representation of gamma functions and then expanding at large $\omega$.

The large $\omega$ behaviour of $G_{LR}(\omega)$ is fixed on general grounds to be\footnote{Firstly, in general $G_{LR}(\omega) = \text{Im}\, G^{tt}_{22}(\omega)/\sinh \frac{\omega}{2T} \sim e^{-\omega/(2T)}\text{Im}\, G^{tt}_{22}(\omega)$. Secondly, at large real $\omega$ we can obtain $\text{Im}\, G^{tt}_{22}(\omega) \sim \omega^{-1}$ directly from the infalling solution $e^{i \omega X(z)}$ in (\ref{eq:hor}). Near the boundary we can use $X(z) \approx z$, expand the exponential $e^{i \omega z}$ in small $z$, and then use the definition (\ref{eq:Gtt}).}
\be\label{eq:known}
G_{LR}(\omega) \; \propto \; \omega^{-1}e^{-\omega/(2T)} \,.
\ee
Matching with (\ref{eq:tomatch}) requires
\be\label{eq:ZZ}
{\mathcal Z}_{\text{I}}= \frac{1}{2T} - 2 \, {\rm Im} \, {\mathcal Z}_{\text{QNM}} \,, \qquad j_0 = \frac{1}{2} \,.
\ee
In the absence of any new set of modes, $\mathcal{Z}_I = 0$. It follows from (\ref{eq:ZZ}) that a single set of QNMs, such as (\ref{eq:analyt}) in the previous section, can only be consistent with the falloff (\ref{eq:known}) if ${\rm Im} \, {\mathcal Z}_{\text{QNM}} = \frac{1}{4 T}$.
We will now explain that the branch cut emanating from the holographic veils obstructs this equality and thereby forces the existence of additional modes. The modes need not be overdamped a priori, although this is the minimal way to satisfy (\ref{eq:ZZ}).

\begin{figure}[h]
    \centering
    \includegraphics[width=0.6\linewidth]{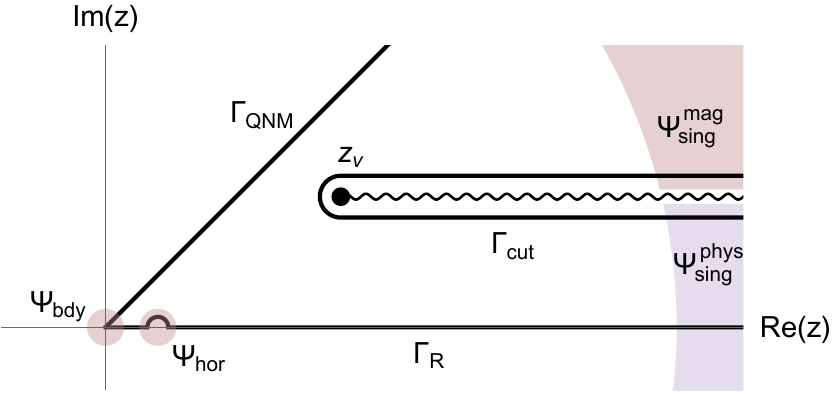}
    \caption{Three contours: $\Gamma_\text{QNM} = \tilde \Gamma_1$ goes to the magnetic singularity, $\Gamma_{\R}$ that runs to the physical mixmaster singularity, looping over the pole at the horizon, and $\Gamma_\text{cut}$ that runs around the branch cut emanating from the holographic veil.}
    \label{fig:gammacut}
\end{figure}

Consider deforming the $\tilde \Gamma_1 \equiv \Gamma_\text{QNM}$ contour down to a contour $\Gamma_{\R}$ along the real axis. As illustrated in Fig.~\ref{fig:gammacut}, this process necessarily picks up an additional contour $\Gamma_\text{cut}$ that winds around the branch cut.
The corresponding integrals of $\frac{1}{h}$ along these contours therefore obey $\mathcal{Z}_\text{QNM} = \mathcal{Z}_{\R} + \mathcal{Z}_\text{cut}$. Now take the imaginary part of this last equality and note that $\text{Im} \, \mathcal{Z}_{\R} = \frac{1}{4 T}$, from the loop around the horizon. This is a standard fact \cite{Fidkowski:2003nf}. Thus we obtain from (\ref{eq:ZZ}) that
\be\label{eq:icut}
{\mathcal Z}_{\text{I}} = - 2 \, \text{Im}\, \mathcal{Z}_\text{cut} \,.
\ee
That is, the decay rates $\gamma_i$ of the overdamped modes in (\ref{eq:gamma}) are controlled by the holographic veil branch cut. Conversely, without the holographic veil (or some other obstruction, such as an inner horizon, as we discuss shortly) there could not be a set of overdamped quasinormal modes. Other black holes with both a family of overdamped modes and a holographic veil are \cite{deOliveira:2024bgh, Oh:2026qcm}.

We have verified that (\ref{eq:icut}) holds on our numerically constructed backgrounds and quasinormal modes. Furthermore in Appendix \ref{app:qnm} we verify that the spacing between modes constructed numerically using only the exterior geometry is correctly captured by (\ref{eq:gamma}). In the future it would be instructive to re-derive (\ref{eq:icut}), along with subleading non-analytic corrections, directly from a WKB computation of the overdamped modes. Similar analyses have been done in \cite{Chen:2026efa, Oh:2026qcm}. Finally, we should note that none of the remaining veils in our mixmaster interior (the first few of which we enumerated in Table \ref{tab:bounces}) occur on the physical sheet. For example, the second magnetic veil is on the second sheet at $z_\text{v} \approx 4.73 \times 10^6 + 5.79 \times 10^5 i$. These further veils, therefore, do not enter the contour deformation and do not result in additional families of overdamped modes.

The above mechanism whereby a holographic veil produces overdamped QNMs is analogous to the way that such modes can be produced by an inner horizon. In those cases the real axis contour must also loop around the inner horizon, and it loops below. Therefore, when $\Gamma_\text{QNM}$ is deformed to the real-axis contour, it picks up an additional loop around the inner horizon. This loop is given by the inverse temperature of the inner horizon, which indeed controls the spacing of the overdamped QNMs of a scalar field in the AdS-Reissner-Nordstr\"om geometry \cite{Dodelson:2025jff, Chen:2026efa}.

There exists an alternative perspective on the connection between an inner horizon, or  the outermost $g_{tt}$ bounce, and overdamped QNMs. It is instructive to compare and contrast with our discussion above. The effective radial potential seen by a particle with large mass $M$ is given by $V(z) = -M^2 g_{tt}(z)$. Stationary points $z_\star$ with $g_{tt}'(z_\star) = 0$ are therefore locations where a particle can `sit'. A trajectory from the boundary with frequency $\omega^2 = - M^2 g_{tt}(z_\star)$ has a turning point at precisely the extremum of the potential. It is well-known that when this happens the divergent density of states at the extremum produces a quasinormal mode. Recalling that $g_{tt} > 0$ in the interior, in the present setup we obtain an overdamped quasinormal mode with decay rate $\Gamma = M \sqrt{g_{tt}(z_\star)}$ \cite{Hartnoll:2020rwq}. Such stationary points must arise whenever there is an inner horizon, as $g_{tt}$ vanishes at both the inner and outer horizons. They also arise, by definition, whenever $g_{tt}$ has a bounce. This argument only applies to the first bounce, as further bounces are hidden behind the potential and inaccessible from the boundary \cite{Hartnoll:2020fhc}. Now, the potential has to do with the function $V$ in the Schr\"odinger equation (\ref{eq:schro}) while the veils we have discussed above are instead captured by $h$ in that equation. The phenomenon described in the present paragraph is therefore logically distinct from our considerations above, even if the conclusions are consistent. The large mass approach naturally gives the lowest overdamped QNM, while in (\ref{eq:icut}) the holographic veil branch cut instead determines the asymptotic spacing of the modes.

\subsection{Singularities in complex boundary time}
\label{sec:timesing}

Null geodesics bouncing off the Schwarzschild singularity lead to singularities in the time-dependent Green's function \cite{Fidkowski:2003nf}, which is the Fourier transform of the two-sided Green's function:
\begin{equation}
 G_{LR}(t)=\int_{-\infty}^{\infty}\frac{d \omega}{2\pi}\,
                  e^{-i \omega t}G_{LR}(\omega) \,.
 \label{eq:timeFourier}
\end{equation}
It was argued in \cite{Dodelson:2025jff} that 
the `bouncing singularities' of $G_{LR}(t)$ can be determined from the thermal product representation of the Green's function combined with leading order QNM asymptotics. In this section we follow the same logic to obtain the singularities in $G_{LR}(t)$ that follow from (\ref{eq:twoproduct}) together with the asymptotics (\ref{eq:wn0}) and (\ref{eq:gamma}).

The bouncing geodesics, in our case, do not bounce off the physical mixmaster singularity but rather off the complex singularity (\ref{eq:magsing}). While complex singularities are not directly physical, they arise as saddle points in the gravitational path integral. They should potentially also be resolved within string theory/quantum gravity and this resolution may then be visible within the QNM spectrum of the dual field theory. It would be interesting to understand whether the stringy and finite $N$ effects discussed in \cite{Komargodski:2026oob, Batra:2026gbq, Buric:2026sit, Dodelson:2026hjm, Liu:2026luf} are also applicable to complex metric singularities.

We will be very brief as the technical steps are identical to the discussion of charged black holes in \cite{Dodelson:2025jff}. Expressing the product (\ref{eq:twoproduct}) in terms of gamma functions and then using a combination of reflection formulae and large $\omega$ limits of the gamma functions gives
\begin{align}
 G_{LR}(\omega)\ & \propto
 \frac{e^{-\omega/(2T)}}{\omega}
 \frac{1}{
  |1- i \sqrt3\,e^{2 i {\mathcal Z}_{\text{QNM}}\omega}|^2 \,
  (1+e^{-(1/T-4\operatorname{Im}{\mathcal Z}_{\text{QNM}})\omega})}  \label{eq:timeExponentialProduct} \\
  & \propto \sum_{r,s,k=0}^\infty  \frac{(i \sqrt3)^r(- i \sqrt3)^s(-1)^k}{\omega}
 \exp\!\left[-\left(k+\frac12\right)\frac{\omega}{T}
       +2 i \omega\bigl((r-k){\mathcal Z}_{\text{QNM}}-(s-k){\mathcal Z}_{\text{QNM}}^*\bigr)\right] \,. \nonumber
\end{align}
Using this final expression in the Fourier transform (\ref{eq:timeFourier}), divergences come from the large $\pm \omega$ regions of the integral. Recalling that $G_{LR}(-\omega) = G_{LR}(\omega)$, the complex-time singularities produced by the approximation (\ref{eq:timeExponentialProduct}) are at
\begin{equation}
 t_{rsk}=\pm\left[\frac{i}{T}\left(k+\frac12\right)
                  +2(r-k){\mathcal Z}_{\text{QNM}}-2(s-k){\mathcal Z}_{\text{QNM}}^*\right] \,,
 \qquad r,s,k\in\mathbb Z_{\geq0} \,.
 \label{eq:timeCandidateSet}
\end{equation}
and the leading singular behaviour is 
\be
 G_{LR;rsk}(t)\big|_{\rm sing}
 =-\frac{K}{2\pi}(i \sqrt3)^r(-i \sqrt3)^s(-1)^k
           \log(t^2-t_{rsk}^2) \,,
\ee
where the overall constant $K$ does not depend on $r,s,k$.

Considering the $+$ modes in (\ref{eq:timeCandidateSet}), the singularities are at
\begin{equation}
 \operatorname{Re}t_{rsk}=2(r-s)\operatorname{Re}{\mathcal Z}_{\text{QNM}},\qquad
 \operatorname{Im}t_{rsk}=\frac{1}{2T}+2(r+s)\operatorname{Im}{\mathcal Z}_{\text{QNM}}
                         +k\left(\frac{1}{T}-4\operatorname{Im}{\mathcal Z}_{\text{QNM}}\right).
 \label{eq:timeCoordinates}
\end{equation}
Both $\operatorname{Im}{\mathcal Z}_{\text{QNM}}$ and $\frac{1}{T}-4\operatorname{Im}{\mathcal Z}_{\text{QNM}}$ are seen to be positive in our numerical backgrounds. Therefore all of these singularities obey ${\rm Im} \, t_{rsk} \geq {1 \over 2 T}$, as required by KMS analyticity (see e.g.~Appendix A of \cite{Dodelson:2025jff}).
On the other hand, in a limit where the sources become small we expect the optical depth ${\mathcal Z}_{\text{QNM}}$ to tend towards its value for Schwarzschild-AdS$_4$. The Schwarzschild-AdS$_4$ singularity is in contradiction with the stronger analyticity argument in \cite{Komargodski:2026oob}. As we noted in \S\ref{sec:abjm}, however, that argument does not obviously apply in the presence of sources.

\section{Entanglement entropy and holographic veils}
\label{sec:entanglement}

We have shown above that BKL behavior in the black hole interior very generally leads to the emergence of holographic veils: extremal surfaces of the transverse area in the complexified geometry. The presence of these surfaces prevents the WKB two-point correlator from probing the real black hole interior. On the other hand, extremal surfaces play a fundamental role in the calculation of holographic entanglement entropy \cite{Ryu:2006bv,Hubeny:2007xt}. In this section we will see that analytic continuation of the entanglement entropy to complex boundary regions, in a large length limit analogous to that in \cite{Hartman:2013qma}, is given by the area of the veils.

Consider a boundary strip ${\cal R}_\ell$ of length $\ell$:
\be
{\cal R}_{\ell} \equiv \left\{ t=0, -\ell/2 < x< \ell/2 \right\} \,.
\ee
The strip is extended along the $y$ direction.
The corresponding bulk Ryu--Takayanagi surface is anchored to the strip ${\cal R}_\ell$ on the boundary and drops into the bulk following a function $z(x)$. To find the bulk surface we need to extremise the area
\be\label{eq:Al}
A(\ell) \equiv L_y \int_{- \ell/2}^{\ell/2} dx \, {\cal L} \,, \qquad {\cal L}={1 \over z^2} \sqrt{1+{e^{-2B(z)} \over f(z)} \left(\frac{dz}{dx}\right)^2} \,.
\ee
Here $L_y$ is a regularisation of the extent of the strip in the $y$ direction.

It will be useful to recall the standard result that arises in taking the real $\ell \to \infty$ limit of the strip \cite{Ryu:2006bv}. In this limit the extremal surface drops straight down to the horizon at $z = z_\mathcal{H}$ and over most of the $x$ range it hugs the horizon with $z'(x) \approx 0$. The Ryu--Takayanagi entanglement entropy in the limit $\ell \to \infty$ is then given by
\be\label{eq:EE}
S_{\text{EE}}( {\cal R}_{\ell})= {A(\ell) \over 4 G_N} = \frac{L_y \ell}{4 G_N} \frac{1}{z_\mathcal{H}^2} + \cdots = s_\text{BH} L_y \ell + \cdots \,.
\ee
Here $s_\text{BH}$ is thermal black hole entropy density. The ellipses include divergent boundary contributions that must be regularised.

In this section we will be considering $A(\ell)$ in (\ref{eq:Al}) as a function of complex $\ell$, and taking large $|\ell|$ limits in the complex plane. We will start with the extremal surface at real positive $\ell$, that describes conventional entanglement entropy, and then analytically continue that saddle point in $\ell$. This is a well-defined process if we take the limit $G_N \to 0$ first, in order to select the saddle, and is analogous to taking the $\Delta \to \infty$ limit for geodesics \cite{Fidkowski:2003nf}. As $|\ell| \to \infty$ along some complex path $\Gamma$, we show below that the surface `sticks' to a complex surface with extremal transverse area density (i.e.~a holographic veil) to give
\be\label{eq:acomplex}
\frac{A_\Gamma(\ell)}{L_y \ell} = \frac{1}{z^2_{\text{v}_\Gamma}} + \cdots \qquad \Rightarrow \qquad \lim_{|\ell|\to \infty} \frac{S_\text{EE}({\cal R}_{\ell})}{L_y \ell} = \frac{{\mathcal A}_{\text{v}_\Gamma}}{4 G_N} \,.
\ee
Here $z_{\text{v}_\Gamma}$ is the location of a holographic veil with area ${\mathcal A}_{\text{v}_\Gamma}$. The mechanism behind (\ref{eq:acomplex}) is the same as that leading to (\ref{eq:EE}), with the veil playing the role of the horizon. The same mechanism also appears in \cite{Hartman:2013qma}, where the linear growth of entanglement in time is controlled by an extremal surface in the black hole interior. Complex extremal surfaces have been discussed previously in e.g.~\cite{Fischetti:2014uxa,Fischetti:2014zja,Heller:2024whi,Heller:2025kvp}. It would be interesting to understand if those works have any connection to our discussion.

The particular holographic veil selected in (\ref{eq:acomplex}) depends on the path $\Gamma$. Indeed, we will shortly show that there are typically branch cuts in $A(\ell)$. We will call these entanglement folds. Taking the contour $\Gamma$ to cross these cuts leads to different veils controlling the large $|\ell|$ limit.

\subsection{Large $\ell$ limit and veils}

The Lagrangian density in (\ref{eq:Al}) has no explicit $x$ dependence. This allows the equation for the extremal surface to be integrated once. Using the solution in (\ref{eq:Al}) gives the extremal area as
\be\label{eq:Astar}
A_\star(\ell) = 2L_y \int_0^{z_0(\ell)}\frac{e^{-B(z)} \, dz}{\sqrt{z^4 f(z) \big[1 - \left(z/z_0(\ell)\right)^4\big]}} \,,
\ee
where the tip $z_0(\ell)$ of the bulk surface is determined by
\be\label{eq:ell}
\ell = 2 \int_0^{z_0(\ell)}\frac{z^2}{z_0(\ell)^2} \frac{e^{-B(z)} \, dz}{\sqrt{f(z) \big[1 - \left(z/z_0(\ell)\right)^4\big]}} \,.
\ee
The factors of 2 here account for the fact that the surface must fall down into the bulk and then rise back up to the boundary. The $z\approx 0$ contribution needs to be regularised in (\ref{eq:Astar}), but this fact has no bearing on our computations below which are concerned with the contribution from $z \approx z_0(\ell)$. To analytically continue (\ref{eq:Astar}) and (\ref{eq:ell}) to complex $\ell$ we change the contour of integration
\be
\int_0^{z_0(\ell)} \to \;\; \int_\Gamma \,.
\ee

Suppose that the path $\Gamma$ takes $z_0(\ell) \to z_\text{v}$, a veil. From our discussion in \S\ref{sec:veil} we know that near a veil $f \approx f_\text{v} (1 - z/z_\text{v})$. Using this form in (\ref{eq:ell}) it is clear that the integral will develop a logarithmic divergence as $z_0(\ell) \to z_\text{v}$, whereupon the two zeros inside the square root coincide. Extracting the divergence one finds,
\be
\ell \approx - \frac{e^{-B_\text{v}} z_\text{v}}{\sqrt{f_\text{v}}} \log \left[1 - \frac{z_{0}(\ell)}{z_\text{v}}\right] \,.
\ee
That is, $|\ell|$ diverges. The extremal area integral (\ref{eq:Astar}) experiences the same divergence and hence
\be
\lim_{|\ell| \to \infty} \frac{A_\star(\ell)}{L_y \ell} = \frac{1}{z_\text{v}^2} \,,
\ee
as we advertised in (\ref{eq:acomplex}). All divergences in $\ell$ arise in this way, from $f$ vanishing, which occurs at veils and horizons. We see in the following \S\ref{sec:ef} that the veil or horizon that is picked out depends on the choice of complex path. By considering suitable $|\ell| \to \infty$ limits of the entanglement entropy in the complex $\ell$ plane, we are able to extract the area of holographic veils.

\subsection{Entanglement folds and an explicit background}
\label{sec:ef}

Firstly, observe that from (\ref{eq:Astar}) and (\ref{eq:ell}) one can show that
\be\label{eq:Aprime}
A_\star'(\ell) = \frac{L_y}{z_0(\ell)^2} \,.
\ee
This fact can be seen by considering $\frac{d}{dz_0}\left(A_\star(z_0) - L_y \ell(z_0)/z_0^2 \right)$. From (\ref{eq:Aprime}) it follows that if $z_0(\ell)$ has a branch cut at some $\ell_\text{b} \equiv \ell(z_\text{b})$, with $z_\text{b} \neq 0$, then so will $A_\star(\ell)$.

\begin{figure}[!t]
    \centering
    \includegraphics[width=\textwidth]{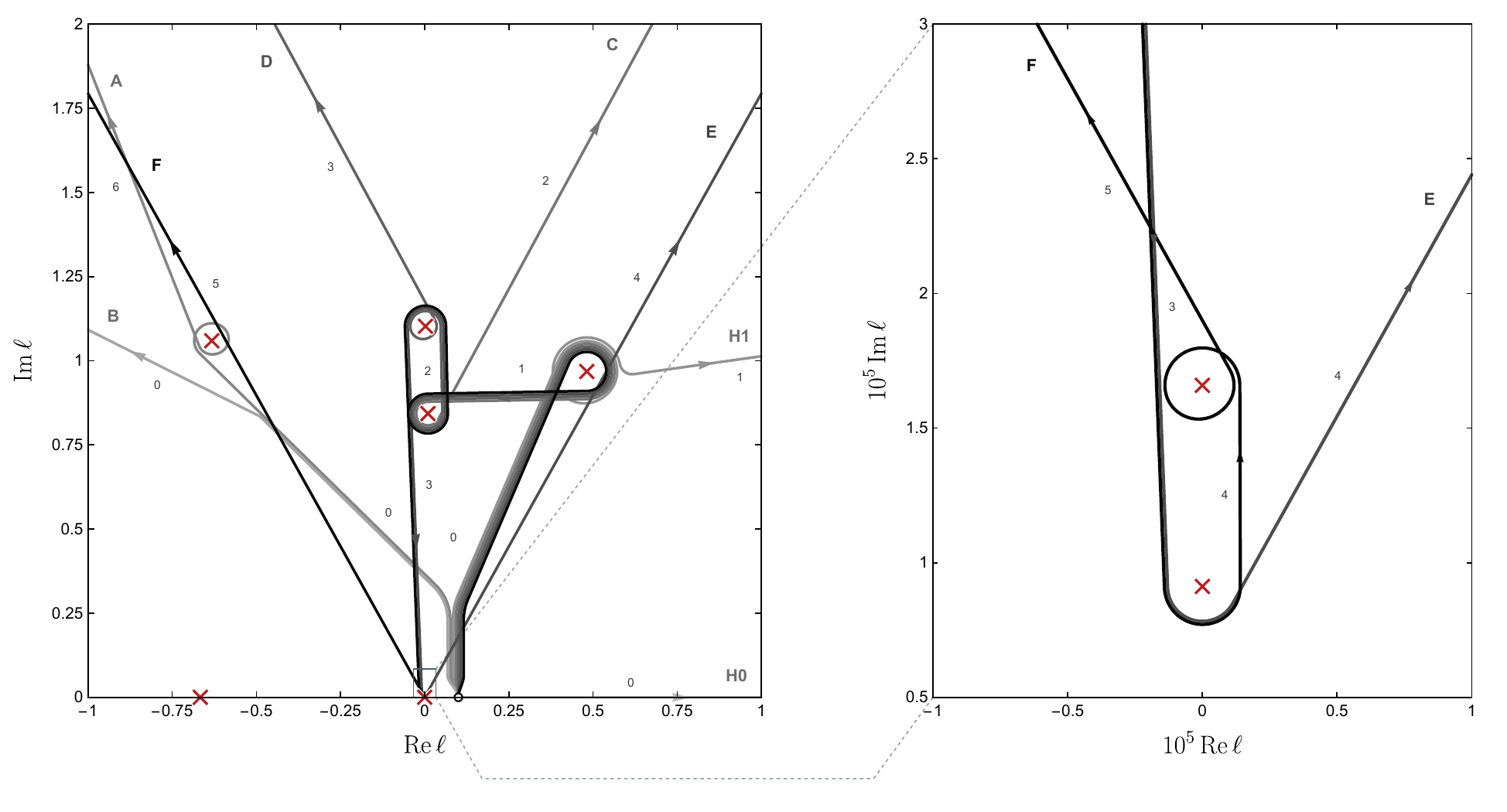}
    \caption{Illustrative contours on the complex $\ell$ Riemann surface, starting at real $\ell=0.1$ (in bulk units where $z_\mathcal{H}=1$). The contours follow different asymptotic angles to large $|\ell|$, and wind around entanglement folds (shown as crosses) onto different sheets. Along the $A,B,C,D,E,F$ contours the asymptotic value of the derivative $A'_\star(\ell)$ is controlled by a holographic veil while for $H0, H1$ it is controlled by the horizon. Each segment of a contour is labelled by its sheet of the $\ell$-plane. The right hand plot is a zoom-in showing the paths around two entanglement folds.
    }
    \label{fig:paths}
\end{figure}

To see how such branch cuts can arise, it is helpful to write (\ref{eq:ell}) as giving $\ell(z_0)$:
\be\label{eq:ellinv}
\ell(z_0) = 2 \int_0^{z_0}\frac{z^2}{z_0^2} \frac{e^{-B(z)} \, dz}{\sqrt{f(z) \big[1 - \left(z/z_0\right)^4\big]}} = 2 \int_0^1 \frac{G(t z_0) \, t dt}{\sqrt{1 - t^4}}\,,
\ee
where in the second step we set $G(z) \equiv z e^{-B(z)}/\sqrt{f(z)}$. A square root branch cut in $z_0(\ell)$ arises at $\ell_\text{b}$ if $\ell'(z_\text{b}) = 0$ with $\ell''(z_\text{b})$ nonzero. That is, whenever
\be
0 = \ell'(z_\text{b}) = 2 \int_0^1 \frac{G'(t z_\text{b}) \, t^2 dt}{\sqrt{1 - t^4}} \,.
\ee
Near such a point,
$z_0(\ell) \approx z_\text{b} \pm [2/\ell''(z_\text{b}) \times(\ell - \ell_\text{b})]^{1/2}$.
From (\ref{eq:Aprime}) we then conclude that
\be
A_\star(\ell) \approx A_{\star \text{b}} + \frac{L_y}{z_\text{b}^2} (\ell - \ell_\text{b}) \mp
\frac{4 L_y}{3 z_\text{b}^3} \sqrt{\frac{2}{\ell''(z_\text{b})}}(\ell - \ell_\text{b})^{3/2}
+ \cdots \,.
\ee
Here we see the branch cut in $A_\star(\ell)$.

In Fig.~\ref{fig:paths} we have followed a series of illustrative curves in the complex $\ell$ plane using, as always, the lower geometry in Fig.~\ref{fig:interior}. Bulk units are taken such that $z_\mathcal{H}=1$. The background is extended numerically to complex $z$. The paths shown all start at real $\ell = 0.1$. We then use (\ref{eq:ellinv}) to obtain $z_0(\ell)$ and hence follow $A_\star'(\ell)$ in (\ref{eq:Aprime}) along the curve. The curves are taken to wind around various entanglement folds. Along the curves we find that $A'_\star(\ell)/L_y \to 1/z^2_\text{v}$ as $|\ell| \to \infty$, where the particular holographic veil $z_\text{v}$ depends on the sheet in the complex $\ell$ plane and the asymptotic angle of the curve:
along the $C,D$ curves the veils are associated to the $M1$ bounce in Table \ref{tab:bounces}, along the $E,F$ curves the veils are associated to the $M2$ bounce in that table, and along the $A,B$ curves the veils are those given in the caption of Table \ref{tab:bounces}, which we have not associated to bounces. Entanglement folds are shown as crosses in the figure. We have observed that several, but not all, of the entanglement folds are also associated to BKL bounces.

The conclusion from the above is that the analytic continuation of the Ryu--Takayanagi entanglement entropy $S_\text{EE}(R_\ell)$ picks out the locations $z_\text{v}$ of holographic veils in the limit $|\ell| \to \infty$.

\section{BKL in thermal one-point functions}
\label{sec:onepoint}

\subsection{Review of the setup}

\begin{figure*}[h]
  \centering
  \includegraphics[width=0.4\linewidth]{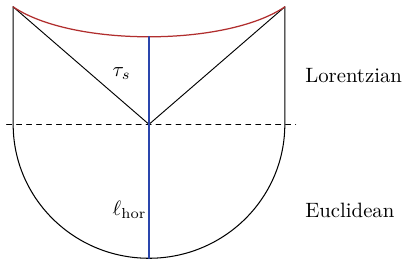}
  \caption{Geodesic that controls the thermal one-point function of the stringy operator $\ocal$.}
  \label{fig:geodesic}
\end{figure*}

This section is concerned with signatures of the BKL interior in the thermal one-point function of a `stringy' operator $\ocal$, dual to a heavy bulk field $\phi$ with complexified mass $m$. Grinberg and Maldacena have shown \cite{Grinberg:2020fdj} that the large-mass behavior of the one-point function takes the form
\be
\langle \ocal\rangle \sim (\text{powers of }m) \,
 e^{-m(\ell_{\rm hor} + i \tau_s)},\qquad \operatorname{Im}m<0 \,.
 \label{eq:GMidea}
\ee
The quantities in this formula are illustrated in Fig.~\ref{fig:geodesic}.
We will review the setup of \cite{Grinberg:2020fdj} and then show how the prefactor of the exponential behaviour (\ref{eq:GMidea}) reveals the interior BKL bounces.

The scalar mode $\phi$ has Euclidean action
\begin{equation}
 \begin{aligned}
 S_{E}={1 \over 2 \kappa^2_4} \int d^4x\sqrt g\,\Big[&\tfrac12(\nabla\phi)^2
       +\tfrac12m^2\phi^2+\alpha\, \phi W^2 + \cdots \Big],
 \end{aligned}
 \label{eq:onepoint}
\end{equation}
where  $W^2 \equiv W_{\mu \nu \rho \sigma} W^{\mu \nu \rho \sigma}$ is the square of the Weyl tensor and $\alpha \lesssim {1 \over \sqrt{\lambda}}$ due to causality \cite{Camanho:2014apa,Meltzer:2017rtf,Cordova:2017zej}. In pure AdS$_4$ we have $W^2=0$ and hence, in the absence of a boundary source, the equation of motion from (\ref{eq:onepoint}) is solved by $\phi=0$.
In the black hole background $W^2 \neq 0$ sources the scalar field in the bulk, leading to a thermal boundary expectation value $\langle \ocal \rangle$ \cite{Myers:2016wsu}. The key observation in \cite{Grinberg:2020fdj} is that the divergence of $W^2$ close to the black hole singularity leaves a signature in this boundary one-point function.

The dimension of the dual operator $\ocal$ can be parametrised as
\be\label{eq:mu}
\Delta = {3 \over 2} - i \mu,
\qquad -\mu^2 \equiv m^2 + {9 \over 4} .
\ee
We will take $\mu \in \mathbb{R}$, which corresponds to an analytic continuation to imaginary mass.
Given that $\Delta \sim \lambda^{1/4}$ for stringy operators, such imaginary scaling dimensions arise naturally upon letting $\lambda \to \lambda e^{- 2 \pi i}$ in the dual CFT.

In \cite{Grinberg:2020fdj} the one-point function is obtained, to classical order in the bulk, as
\be
\langle \ocal \rangle = c(\mu) \ocal(\mu) \,,
\ee
where $c(\mu)$ is a constant\footnote{In a normalisation where $\langle {\cal O}(x) {\cal O}(0) \rangle = x^{-2 \Delta}$ the constant is $c(\mu)^2 = - \a^2 \pi^{3/2} \Gamma(\Delta - \frac{1}{2})/(4 \mu^2\k_4^2 \Gamma(\Delta))$.} and
(working in bulk units with $z_\mathcal{H}=1$)
\begin{equation}
\ocal(\mu) \equiv -\int_0^1\frac{s(z) W^2(z)}{z^4}G_\ocal (z)\,dz \,.
 \label{eq:euclidean}
\end{equation}
Here $G_{\ocal}(z)$ solves
\be\label{eq:wave}
- \frac{z^4}{s} \frac{d}{dz} \left(\frac{sf}{z^2} \frac{dG_\ocal}{dz}\right) + m^2 G_\ocal = 0 \,,
\ee
is regular at the Euclidean horizon and is normalised so that $G_\ocal(z) \approx z^{3-\Delta}$ as $z \to 0$, corresponding to a unit boundary source. Note that $G_\ocal(z)$ will also have a `normalisable' $z^\Delta$ mode as $z \to 0$. It may be confusing that a boundary source appears in the computation of an expectation value generated from the bulk alone. This boundary condition arises by considering the response at some finite bulk point and then taking the limit to the boundary \cite{Grinberg:2020fdj}.

Our task is to understand the integral \eqref{eq:euclidean} at large $\mu$ in our $M$-theory black hole geometry. The integral converges absolutely for $-3<{\rm Re}(\Delta)<6$, which includes our $\Delta = \frac{3}{2} - i \mu$. In \cite{Grinberg:2020fdj} the integral was evaluated to leading exponential order to give (\ref{eq:GMidea}). Below we obtain the prefactor of the exponentials in (\ref{eq:GMidea}). To that end we consider the ratio
\be
F(\mu) \equiv {\ocal(\mu) \over \frac{16\pi^{3/2}}{3^{5/2}} (- i \mu)^{5/2} e^{i \mu (\ell_{\rm hor}+i\tau_s)}} \,.
\label{eq:normalized1pt}
\ee
The normalisation in \eqref{eq:normalized1pt} has been chosen such that $F(\mu) \to 1$ at large $\mu$ in Schwarzschild-AdS$_4$. The power of $\mu^{5/2}$ will be reviewed below. The factors of $\ell_\text{hor}$ and $\tau_s$ belong to the background being studied (not Schwarzschild in general) in order to cancel the leading behaviour (\ref{eq:GMidea}).

We will show that the pre-factor $F(\mu)$ has zeros and maxima associated with BKL bounces and whose locations contain information about the Kasner exponents. This observation, however, has an important caveat:
higher-derivative terms that were dropped in the action \eqref{eq:onepoint} are expected to contribute to 
$F(\mu)$ at the same order as our results from the Weyl-squared term, cf.~footnote $6$ in \cite{Grinberg:2020fdj}. Therefore the analysis in the present section does not correspond to a controlled calculation of stringy one-point functions in the planar ABJM theory for some given operator. It would be very interesting to understand if these effects can be systematically taken into account, thereby embedding the BKL signatures we uncover below into ABJM thermal one-point functions.

\subsection{Proper length coordinate and deformation of the contour}

It is convenient, following \cite{Grinberg:2020fdj}, to introduce the proper radial distance to the horizon
\begin{equation}
 \rho(z) \equiv \int_z^1\frac{dw}{w\sqrt{f(w)}} \,.
 \label{eq:rho}
\end{equation}
The horizon is at $\rho = 0$ and the AdS boundary at $\rho \to \infty$. The integral (\ref{eq:euclidean}) becomes
\be\label{eq:int2}
\ocal(\mu) = -\int_0^\infty  v(\rho) W^2(\rho) G_\ocal(\rho)\,d\rho \,,
\ee
where we introduced the volume density of the constant $\rho$ slices
\be\label{eq:vv}
v(z) \equiv \sqrt{-g_{tt} g_{xx} g_{yy}} = \frac{\sqrt{f(z)} s(z) }{z^3} \,.
\ee
The wave equation (\ref{eq:wave}) is now
\be\label{eq:wave2}
 \ddot G_\ocal + \frac{\dot v}{v} \dot G_\ocal = m^2 G_\ocal \,,
\ee
where dots are derivatives with respect to $\rho$. Let us note already that $v \sim \sqrt{f s^2} = \sqrt{hs}$ is regular at veils. Holographic veils will play no role in this section.

To analyse the integral \eqref{eq:int2} at large imaginary $\Delta$ we will deform the integration contour. The first step is to extend the integration range to go from $-\infty$ to $+\infty$. Close to the horizon we have $\rho(z) \sim \sqrt{1-z}$. We can go all the way around the horizon by letting $(1-z) \to e^{-2 \pi i} (1-z)$, which corresponds to $\rho \to e^{-i \pi} \rho = - \rho$. The background is analytic at the horizon, which means that it has an expansion in positive powers of $(1-z) \sim \rho^2$. Thus the background is symmetric under $\rho \to - \rho$. However $v$ in (\ref{eq:vv}) will be odd  under $\rho \to - \rho$ because of the $\sqrt{f}$ term. Using these facts we now explain how to define a second solution $\psi_\ocal(\rho)$ to the wave equation (\ref{eq:wave2}) such that the integral (\ref{eq:int2})
\be \label{eq:propercontour}
\ocal(\mu) = \int_{-\infty - i 0}^{\infty-i0}  v(\rho) W^2(\rho) \psi_\ocal(\rho)\,d\rho \,,
\ee
where the integration contour runs just below the real axis. 
Given (\ref{eq:propercontour}) we can deform the contour into the lower half plane. The black hole interior is at $\rho = - i \tau$. A dominant contribution to the integral is found by deforming the contour close to the singularity, where the Weyl curvature becomes large.

We obtain (\ref{eq:int2}) from (\ref{eq:propercontour}) if
\be\label{eq:disc}
\psi_\ocal(\rho) - \psi_\ocal(- \rho - i 0) = - G_\ocal(\rho) \,. 
\ee
Note that $\psi_\ocal(\rho)$ and $\psi_\ocal(- \rho - i 0)$ will both separately solve the wave equation (\ref{eq:wave2}), from the oddness of $v(\rho)$. We therefore only need to specify the solution $\psi_\ocal(\rho)$. We need not impose that $\psi_\ocal(\rho)$ be regular on the horizon, because any singular behaviour $\psi_\ocal(\rho) \sim \log \rho$ is canceled out by the difference in (\ref{eq:disc}), so that $G_\ocal(\rho)$ is regular at the horizon as required. We are therefore free to impose the additional condition that $\psi_\ocal(\rho)$ has  pure source behaviour at the boundary. This means that $\psi_\ocal(\rho) \sim e^{- i \mu \rho}$ at large positive $\rho$, because $\rho \sim - \log z$ as $z \to 0$ from (\ref{eq:rho}), with no $e^{+ i \mu \rho}$ term. While $\psi_\ocal(\rho)$ has sign-definite oscillations as $\rho \to +\infty$, the other sign will be generated for $G_\ocal(\rho)$ by the $\psi_\ocal(-\rho - i 0)$ term in (\ref{eq:disc}). Finally, we must normalise $\psi_\ocal(\rho)$ such that (\ref{eq:disc}) gives the correct unit-source normalisation of 
$G_\ocal(\rho)$.

\paragraph{Schwarzschild example:} For the Schwarzschild-AdS$_4$ background,
\be\label{eq:ss}
W(z)^2 = 12 z^6, \qquad v(z) = {\sqrt{1-z^3} \over z^3} \,, \qquad z^3= {1 \over \cosh^2 \left( {3 \rho \over 2} \right) } \,.
\ee
The black hole singularity is located at $\rho = - i \pi/3 \equiv - i \tau_s$. The regularised distance to the horizon is $\ell_\text{hor} \equiv \lim_{\rho \to \infty} [\rho + \log z] = \frac{\log 4}{3}$.
The wave equation (\ref{eq:wave}) can be solved to give
\be
G_{\ocal}(z) = \frac{\Gamma\!\left(\frac12-\frac{i\mu}{3}\right)^2}
            {\Gamma\!\left(-\frac{2i\mu}{3}\right)}
       z^{\frac32-i\mu}\,
       {}_2F_1\!\left(
       \frac12-\frac{i\mu}{3},
       \frac12-\frac{i\mu}{3};
       1;1-z^3\right) \ . 
\ee
Asymptotically $G_{\ocal}(z) \approx z^{\frac{3}{2} + i 
\mu} + b \, z^{\frac{3}{2} - i \mu}$, with $b$ a constant.
The integral (\ref{eq:euclidean}) is exactly
\begin{align}
\ocal(\mu) &=\frac{8\pi i\mu}{27}\left(\mu^2+\frac94\right)
 \frac{\Gamma(\tfrac12-\tfrac{ i\mu}{3})^2}
 {\Gamma(1-\tfrac{2 i\mu}{3})\cosh(\tfrac{\pi\mu}{3})} \nonumber \\
 &=\frac{16\pi^{3/2}}{3^{5/2}}(-i \mu)^{5/2} e^{i \mu (\ell_{\rm hor}+ i \tau_s)}
 \left[1-\frac{3 i}{8\mu}+O(\mu^{-2})\right].   \label{eq:ifin}
\end{align}
In the final expression we obtained the normalisation factor used in (\ref{eq:normalized1pt}).

We will shortly reproduce the result (\ref{eq:ifin}) using a large $\mu$ analysis along a deformed contour. Firstly, however, note that the second solution to the wave equation is
\be
\psi_{\ocal}(z)
= -\tanh\!\left(\frac{\pi\mu}{3}\right)
  z^{\frac32+i\mu}\,
  {}_2F_1\!\left(
  \frac12+\frac{i\mu}{3},
  \frac12+\frac{i\mu}{3};
  1+\frac{2i\mu}{3};
  z^3
  \right) \,.
\ee
Asymptotically now $\psi_{\ocal}(z) \approx - \tanh \frac{\pi \mu}{3} z^{\frac{3}{2}+ i \mu} + \cdots$ with no $z^{\frac{3}{2} - i \mu}$ term. One can verify (\ref{eq:disc}), being careful with the branch cut at $\rho=0$. If we write $\psi_\ocal(z) \equiv - (1 - A(\mu)) z^{\frac{3}{2}+ i \mu} + \cdots$ then expanding the $\tanh$ at large $\mu$ gives an exponentially small correction
\be\label{eq:Asmall}
A(\mu) \sim 2 e^{- {2 \pi \mu \over 3}} \qquad \Rightarrow \qquad \psi_\ocal(z) \approx - z^{\frac{3}{2}+ i \mu} \,.
\ee
The final relation in (\ref{eq:Asmall}) is important and holds generally, see Appendix \ref{app:normone}. In terms of the proper distance, (\ref{eq:Asmall})
implies that in the large $\mu$ WKB approximation and at large positive $\rho$,
\be
\label{eq:wkbschw}
\psi_{\ocal}(\rho)= - e^{i \mu \ell_{\text{hor}}} {e^{- i \mu \rho} \over \sqrt{v}} \ . 
\ee
The factor of $\ell_{\text{hor}}$ arises from the definition just below (\ref{eq:ss}). From Appendix \ref{app:normone}, (\ref{eq:wkbschw}) holds in all backgrounds, not just Schwarzschild-AdS.

The Euclidean integral (\ref{eq:euclidean}) is highly oscillatory. It is more convenient to use the representation (\ref{eq:propercontour}), which can be deformed into the lower half plane where the WKB form (\ref{eq:wkbschw}) for $\psi_\ocal(\rho)$ gives an exponential suppression. We cannot deform the entire contour below the singularity. Furthermore, the WKB form breaks down as the contour comes close to the singularity because the volume $v$ in (\ref{eq:vv}) vanishes at the singularity and hence the potential $\frac{\dot v}{v}$ in (\ref{eq:wave2}) diverges. We can, however, match (\ref{eq:wkbschw}) onto a local solution near the singularity. In that regime we can use the fact that $v$ has a simple zero at the singularity to write
\be
v(\rho) \approx  v_s (\rho + i \tau_s) \,.
\ee
Recall that $\tau_s$ was defined below (\ref{eq:ss}).
The wave equation (\ref{eq:wave2}) is then, at large $\mu$,
\be
\ddot \psi_\ocal + \frac{1}{\rho + i \tau_s} \dot \psi_\ocal + \mu^2 \psi_\ocal = 0 \,.
\ee
The solution to this equation that matches onto (\ref{eq:wkbschw}) at large $\rho$ is 
\be\label{eq:nearsing}
\psi_{\ocal}(\rho) =
- e^{i \mu (\ell_{\text{hor}} + i \tau_s)} \sqrt{\frac{\pi\mu}{2v_s}}\,
e^{-i\pi/4}
H_0^{(2)}\!\big(\mu(\rho+i\tau_s)\big) \,.
\ee
For Schwarzschild we have $1/\sqrt{2v_s} = i/\sqrt{3}$ and recall also that $\tau_s = \frac{\pi}{3}$.\footnote{The WKB solution on the $\rho \to - \infty$ side of the matching regime (\ref{eq:nearsing}) is
\be
\psi_\ocal(\rho) =
  -\frac{e^{i\mu\ell_{\mathrm{hor}}}}{\sqrt{v}}
  \left(
    e^{-i\mu\rho}
    + 2 i e^{-2\mu\tau_s}e^{i\mu\rho}
  \right) \,.
\ee
Importantly, the $e^{i\mu\rho}$ term generated here is smaller than the first term for $\text{Im}\rho > -\tau_s$.
}

Near the singularity, the integrand of (\ref{eq:propercontour}) becomes large
\be\label{eq:scaled-measure}
v W^2 \approx - \frac{32}{9 (\rho + i \tau_s)^3} \,.
\ee
Because of this growth, the region $|\rho+i\tau_s|=O(1/\mu)$ captures the leading $\mu^{5/2}$ contribution in \eqref{eq:ifin}. Using (\ref{eq:nearsing}) and (\ref{eq:scaled-measure}) and setting $s= \mu(\rho+i\tau_s)$ in (\ref{eq:propercontour}), we get
\be\label{eq:i12}
\ocal(\mu) \approx \frac{32}{9}\,\sqrt{\frac{\pi\mu}{3}}\,
e^{i\mu(\ell_{\mathrm{hor}}+ i \tau_s)+i\pi/4}\,\mu^2
\int_{\mathcal C}\frac{H_0^{(2)}(s)}{s^3}\,ds.
\ee
The contour $\mathcal{C}$ passes above the singularity. The integral in (\ref{eq:i12}) is easily evaluated by residues after noticing that $H_0^{(2)}=2J_0-H_0^{(1)}$ and that the $H_0^{(1)}$ part evaluates to zero by deforming the contour to the upper half-plane. We are then left with $2\int_{\mathcal C} (J_0(s)/s^3) \, ds = {i \pi \over 2}$. This last integral can be evaluated by residues using the fact that $J_0(-s) = J_0(s)$. Thus we obtain
\be
\label{eq:Schw1pt}
\ocal(\mu) \approx \frac{16\pi^{3/2}}{3^{5/2}} (-i \mu)^{5/2} e^{i \mu (\ell_{\rm hor}+i\tau_s)} \,,
\ee
in agreement with (\ref{eq:ifin}).

We will now see that the above steps generalise to other black hole interiors, with the assumption that the first non-analyticity encountered by $v W^2$ in the lower-half plane is the singularity at $\rho = - i \tau_s$. A very simple generalisation of the computations we have just done is the case with a single Kasner epoch near the singularity.

\paragraph{Kasner example:} A near-singularity Kasner metric can be written
\begin{equation}
 d s^2=-d\tau^2+\sum_{j}a_j^2(\tau_s-\tau)^{2p_j}d x_j^2\,,
 \qquad \sum_{j} p_j = 1 \,.\label{eq:Kasner-metric}
\end{equation}
Here $\rho = - i \tau$ and the singularity is at $\tau=\tau_s$. We remain in four dimensions. From (\ref{eq:Kasner-metric}),
\be
\label{eq:kasnercase}
W^2=\frac{k}{(\rho+ i \tau_s)^4} \,, \qquad v=-a_t a_x a_y (\rho+ i \tau_s) \ \,,
\ee
where $k \equiv \frac{4}{3}\left(1-\sum_i p_i^2\right)^2-16p_tp_xp_y$.
For a vacuum Kasner solution $\sum_i p_i^2=1$, but this need not be the case more generally. For example, for the Einstein-scalar Kasner solutions considered in \cite{Hartnoll:2026vhu} this condition does not hold. The only difference from the Schwarzschild background discussed above is the value of the coefficients. The matching and evaluation of the integral proceed in exactly the same way to give
\be
\label{eq:kasneronept}
\ocal(\mu) \approx \frac{k \, \pi^{3/2}}{2\sqrt2} \sqrt{a_t a_x a_y} (-i \mu)^{5/2} e^{i \mu (\ell_{\text{hor}}+i \tau_{s})} \,.
\ee
The Schwarzschild result \eqref{eq:Schw1pt} is recovered setting $p_t = -\frac{1}{3}$, $p_x = p_y = \frac{2}{3}$, $a_t = (\frac{3}{2})^{-1/3}$ and $a_x=a_y = (\frac{3}{2})^{2/3}$. We have also checked numerically that (\ref{eq:kasneronept}) holds for the backgrounds $\mathtt{EinSV}_1$ and $\mathtt{EinSV}_2$ considered in \cite{Hartnoll:2026vhu}. For example, for $\mathtt{EinSV}_1$  we have $(p_t,p_x,p_y)=\left(-\frac17,\frac47,\frac47\right)$, $(a_t,a_x,a_y)\approx \left(0.84, 1.42, 1.42\right)$ and
$\ell_{\text{hor}} - i \tau_s \approx 0.2795 - i 0.6943$.

\subsection{An ideal bounce leads to zeros in the one-point function}
\label{sec:imprint}

A more nontrivial and interesting step is to obtain the imprint of a bounce in the black hole interior on the one-point function. We will use the magnetic bounce solution from \S\ref{sec:mbounce}. The coordinates used by BKL \cite{Belinsky:1970ew} are convenient here. The metric takes the form
\be\label{eq:gga}
d s^2 = -d\tau^2+e^{2\beta_t}d t^2+e^{2\beta_x}d x^2+e^{2\beta_y}d y^2 \,.
\ee
A time coordinate $- \infty < \sigma < \infty$ is introduced such that
\be
\frac{d\tau}{d\sigma} ={e^{- \left( u + {1 \over u} \right)\sigma}\cosh\sigma \over \mu_0} \,.
\ee
Here $u$ and $\mu_0$ are constants. The bounce is then
\be
 e^{\beta_t} =\frac{a_t}{\cosh\sigma}, \qquad e^{\beta_x}=a_xe^{-u\sigma}\cosh\sigma,\qquad
 e^{\beta_y}=a_ye^{-\sigma/u}\cosh\sigma \,.
\label{eq:wall-metric}
\ee
These coordinates relate to those in \S\ref{sec:mbounce} by 
$\frac{z}{z_{\rm m}}=e^{k\sigma}\operatorname{sech}\sigma$, with
$z_{\rm m}=\frac1{\sqrt{a_xa_y}}$ and $k=\frac12\left(u+\frac1u\right)$.
The bounce solution was first obtained in
\cite{rosen1962symmetries} in the context of Einstein-Maxwell theory. In 4d Einstein-Maxwell theory, electromagnetic duality preserves the electromagnetic stress tensor in the Einstein equations. Therefore, the analysis of this section also applies to the electric wall bounces in the M-theory mixmaster.

A benefit of having the geometry in the form (\ref{eq:gga}) and (\ref{eq:wall-metric}) is that it makes manifest that the bounce describes a transition between a pair of Kasner epochs (\ref{eq:Kasner-metric}). The epochs before and after the bounce are parameterised by the Lifshitz-Khalatnikov parameter $u>1$,
\begin{equation}
 \begin{aligned}
 \text{before:}\qquad
 (p_t^-,p_x^-,p_y^-)&=\frac{(-u,\ u(u+1),\ u+1)}{u^2+u+1} \,,\\[2pt]
 \text{after:}\qquad
 (p_t^+,p_x^+,p_y^+)&=\frac{(u,\ u(u-1),\ 1-u)}{u^2-u+1} \,.
 \end{aligned}\label{eq:K-u}
\end{equation}
These satisfy $\sum_i p_i = \sum_i p_i^2 = 1$. Before the bounce, space expands along the $t$ direction and contracts along $x$ and $y$. After the bounce the $y$ direction expands while $t$ and $x$ contract.

From the background above we obtain the quantities needed to calculate the one-point function:
\begin{equation}
\begin{aligned}\label{eq:Wv2}
W^2 &=
  16\mu_0^4
  \frac{e^{4\left(u+\frac{1}{u}\right)\sigma}\tanh^2\sigma}
       {\cosh^4\sigma}
  \left(
    \left(u+\frac{1}{u}\right)^2 - 1
    - 3\left(u+\frac{1}{u}\right)\tanh\sigma
    + 3\tanh^2\sigma
  \right), \\
v &=
  -i a_t a_x a_y\,e^{-\left(u+\frac{1}{u}\right)\sigma}
  \cosh\sigma \,.
\end{aligned}
\end{equation}
Here $v$ was defined in (\ref{eq:vv}), and is imaginary in the interior.
The square of the Weyl tensor diverges as we approach the singularity as $\sigma \to \infty$, as expected. Remarkably, it also exhibits a double zero at the location of the wall $\sigma =0$.
This will play an important role in what follows. 

We next calculate $\ocal(\mu)$ in the background (\ref{eq:wall-metric}). As a function of large $\mu \gg 1$, the one-point function transitions between the Kasner expressions \eqref{eq:kasneronept}, which we will write as $\ocal_{\pm}(\mu)$. For $\mu \ll \mu_0$
the Kasner exponents are $\{p_i^-\}$ and for $\mu_0 \ll \mu$ the Kasner exponents are $\{p_i^+\}$. There is a nontrivial crossover between these two limits that we proceed to compute. At large $\mu$ the wave equation is
\begin{equation}
 \frac{d^2\psi_\ocal}{d \sigma^2}
 -\left(\frac{\mu}{\mu_0}\right)^2e^{-2(u+{1 \over u})\sigma}\cosh^2\sigma\,\psi_\ocal=0 \,,
 \label{eq:wall-wave}
\end{equation}
and we again need the solution that matches onto (\ref{eq:wkbschw}). Recalling that $\rho = - i \tau$, this becomes the requirement that as $\sigma \to - \infty$,
\be\label{eq:wallBC}
\psi_\ocal(\sigma) \approx -\frac{e^{i\pi/4+i \mu(\ell_{\text{hor}} +i \tau(\sigma))}}{\sqrt{a_ta_xa_y\,e^{-(u+{1 \over u}) \sigma}\cosh \sigma}} \,.
\ee
The full solution to (\ref{eq:wall-wave}) can then be used in the integral (\ref{eq:propercontour}),
\be
\label{eq:wallint}
\ocal(\mu) = \int_{\mathcal C} v(\rho) W^2(\rho) \psi_\ocal(\rho) d\rho \,.
\ee
To extract the dominant contribution we can take ${\mathcal C}$ to be the line $\rho = \frac{x}{\mu} - i \tau_c$, with $x \in (-\infty,\infty)$ and $\tau_c < \tau_s$ arbitrary. Due to the analyticity of the integrand the result does not depend on $\tau_c$. 

\begin{figure*}[t]
  \centering
  \includegraphics[width=1\linewidth]{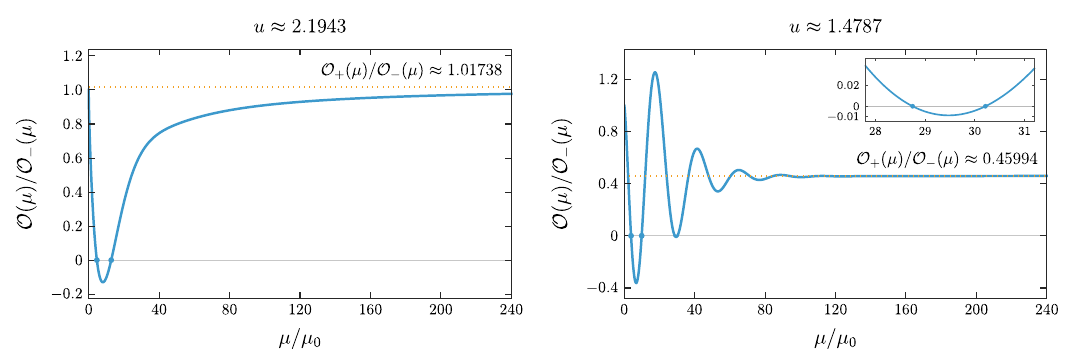}
  \caption{Two examples of the mass dependence of the thermal one-point function $\ocal(\mu)$ for an isolated interior bounce, normalised to the Kasner value before the bounce $\ocal_-(\mu)$. The left plot has
  $u \approx 2.19$ and shows two zeros. The right plot has $u\approx 1.48$
    and has four zeros. The dotted lines mark the
    large-$\mu$ limits $\ocal_+(\mu)/\ocal_-(\mu)$.}
  \label{fig:rosen-wall-onepoint-panels}
\end{figure*}

The differential equation (\ref{eq:wall-wave}) can be solved numerically and then the integral (\ref{eq:wallint}) also calculated numerically. In this way we have plotted $\ocal(\mu)$ as a function of ${\mu \over \mu_0}$ for two illustrative bounces in Fig.~\ref{fig:rosen-wall-onepoint-panels}.
In the figure we verify that for $\mu \gg \mu_0$, it is indeed the case that $\ocal(\mu)$ approaches the inner Kasner value of $\ocal_+(\mu)$. Thus we can think of $\ocal(\mu)$ as tracking the interior geometry. A striking aspect of the plots are the zeros of $\ocal(\mu)$. We will now elaborate on these.

The number of zeros depends on the value of the parameter $u$. A pair of zeros is a robust feature, with two additional zeros appearing for some range of $u$. The presence of at least two zeros is a consequence of the double zero of $W^2$ that we noted in (\ref{eq:Wv2}). This can be seen as follows. The wave equation (\ref{eq:wall-wave}) describes a scattering process in which the asymptotic wave (\ref{eq:wallBC}) is the first order term in a Born series solution. This is a convergent series and the first order term is found, numerically, to give an excellent approximation to the full integral in (\ref{eq:wallint}). With this approximation, (\ref{eq:wallint}) is a Fourier transform that we can invert. Let $F_0(\mu)$ denote the
normalized one-point function \eqref{eq:normalized1pt} obtained by using
the incident wave \eqref{eq:wallBC} throughout the integral
\eqref{eq:wallint}. For the isolated bounce, the continuation of $\sqrt{v}\,W^2$ from the real interior is analytic in the half-plane above the singularity, $\operatorname{Re}(\tau_s-i\rho)>0$, and decays at infinity. The inverse transform, evaluated on the interior at $\rho = - i \tau$, therefore gives
\be\label{eq:inv}
  \sqrt{i\,v(-i\tau)}\,W^2(-i\tau)
  = \frac{8\sqrt{\pi}}{3^{5/2}}
    \int_0^\infty \mathrm{d}\mu\,
    \mu^{5/2} F_0(\mu)\,e^{-\mu(\tau_s-\tau)} \,.
\ee
Evaluating (\ref{eq:inv}) and its derivative at the value $\tau_o$ where $W^2(\tau_o) = 0$ gives two integrals of $F_0(\mu)$ that must vanish. Because $F_0(\mu)$ is real for real $\mu$, these conditions require $F_0(\mu)$ to have at least two sign changes, and hence zeros. More generally, (\ref{eq:inv}) is a useful equation that expresses how $\mu$ acts as a `clock' in the interior, conjugate to evolution in proper time to the singularity.

\begin{figure}[h]
  \centering
  \includegraphics[width=\linewidth,height=6.4cm,keepaspectratio]{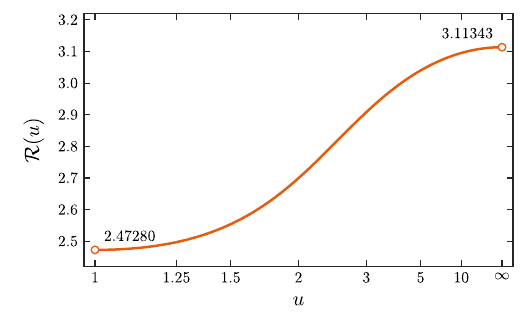}
  \caption{Ratio (\ref{eq:ratioread}) of locations of the first two positive zeros in $\ocal(\mu)$ for an ideal bounce, as a function of the bounce parameter $u$. Open circles mark the limits $u\to1$ and $u\to \infty$.}
  \label{fig:rosen-wall-zero-ratio}
\end{figure}

Denote the locations of the first two zeros as $\mu_{1}^{(0)}(u,\mu_0)$ and $\mu_{2}^{(0)}(u,\mu_0)$. Because $\mu_0$ just sets the scale for $\mu$ in the wave equation (\ref{eq:wall-wave}), the ratio of these two zeros can only be a function of $u$. The ideal bounce geometry therefore gives a universal function that we parameterise as
\be
\label{eq:ratioread}
{\cal R}(u) \equiv R\left(\mu_1^{(0)}(u,\mu_0),\mu_2^{(0)}(u,\mu_0)\right) \,, \qquad R(\mu_1,\mu_2) \equiv
\frac{\sqrt{\mu_{2}^{2} + \frac{9}{4}}}{\sqrt{\mu_{1}^{2} + \frac{9}{4}}} \ . 
\ee
In the second expression here, for improved comparison with numerical finite $\mu$ results later, we have restored the factor of $\frac{9}{4}$ in (\ref{eq:mu}). We dropped this subleading shift in the wave equation (\ref{eq:wall-wave}). To restore the shift in the wave equation we set $\mu^2 \to \m^2 + \frac{9}{4} = -m^2$ before solving. Repeating the numerical process that led to Fig.~\ref{fig:rosen-wall-onepoint-panels} for all $u$ we have generated a plot of ${\cal R}(u)$ in Fig.~\ref{fig:rosen-wall-zero-ratio}.
The curve in Fig.~\ref{fig:rosen-wall-zero-ratio} is monotonic on the ideal branch shown. In the following section we will use this fact to extract information from a plot of $\ocal(\mu)$ obtained from our M-theory mixmaster interior. Namely, from the locations of the zeros of $\ocal(\mu)$ we will be able
to read off the bounce parameter $u$, and hence the Kasner exponents on each side of the bounce.

\subsection{Mixmaster chaos in the thermal one-point function}

The ingredients are now in place to consider the one-point function in our M-theory black hole. Below (\ref{eq:inv}) we observed that $\mu$ is a clock in the interior that is approximately conjugate to the proper distance $\tau$ to the singularity. The expectation, then, is that each bounce in the interior will lead to a pair (at least) of zeroes in $\ocal(\mu)$. The full function is therefore expected to exhibit a nontrivial pattern of zeros.

\begin{figure}[h]
  \centering \includegraphics[width=\linewidth,height=7cm,keepaspectratio]{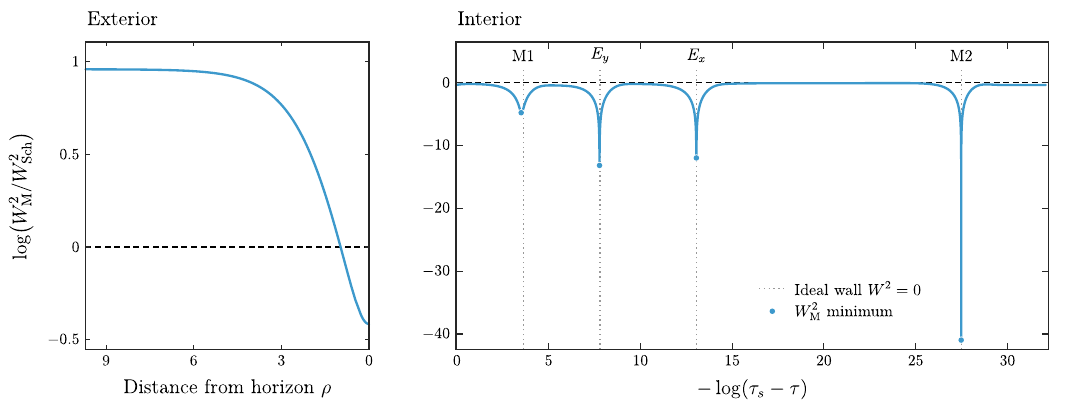}
  \caption{Logarithm of Weyl square $W_{\text{M}}^2$ in the M-theory mixmaster, over the Schwarzschild value $W^2_\text{Sch}$ with the same entropy density. Left plot shows the black hole exterior and right, the interior. The background is the same as the billiard trajectory in the right hand panel of Fig.~\ref{fig:mixmaster}. By comparing to Fig.~\ref{fig:mixmaster} we can identify the four bounces visible in the plot above as a magnetic bounce $M1$ (off the lower wall), followed by an electric $E_y$ bounce (off the left wall), an electric $E_x$ bounce (off the right wall) and finally a second magnetic bounce $M2$ off the bottom wall.}
  \label{fig:weylmixmaster}
\end{figure}

Fig.~\ref{fig:weylmixmaster} shows the logarithm of Weyl 
squared for our M-theory black hole as a function of exterior radial coordinate and interior time.
In the exterior, $W^2$ is featureless relative to the Schwarzschild exterior. In the interior, however, the mixmaster geometry exhibits
sharp spikes. These are points where $W^2$ is becoming very small, as predicted by the double zeros in the ideal bounce solution (\ref{eq:Wv2}).  
The overall curvature is, of course, increasing as the singularity is approached; that is why the near-singularity regime is able to influence the one-point function. It may appear counterintuitive, however, that the bounces themselves manifest as having a small Weyl squared relative to the overall increasing curvature.

\begin{figure*}[h]
  \centering
  \includegraphics[width=\textwidth]{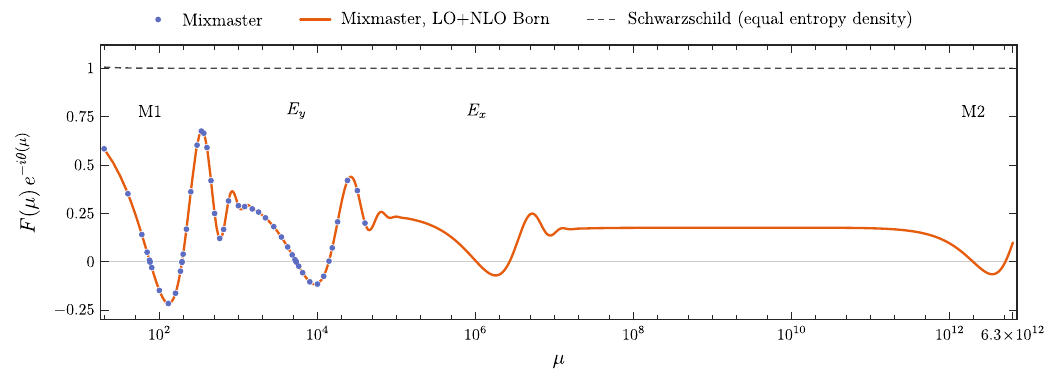}
  \caption{Normalised thermal one-point function \eqref{eq:normalized1pt} in the mixmaster background of Fig.~\ref{fig:weylmixmaster}, stripped of its trivial phase $e^{i \theta(\mu)} \equiv e^{i ({\pi \over 4} - \mu \ell_{\text{hor}})} {G_{\ocal}(1) \over |G_{\ocal}(1)|}$. The blue dots are exact numerical results. The orange curve is computed using a deformed horizontal contour, with the wave equation solved in a Born series approximation as described in the text. The plot exhibits characteristic pairs of zeros. These encode the Kasner exponents on either side of the corresponding bounces ($M1$, $E_y$, $E_x$ and $M2$) in the interior.}
  \label{fig:short-wall-entry-slope}
\end{figure*}

With the Weyl squared at hand, we can calculate and plot the normalised one-point function $F(\mu)$ in \eqref{eq:normalized1pt} as a function of $\mu$. Recall that $F(\mu)$ is equal to $\ocal(\mu)$ relative to its Schwarzschild-AdS$_4$ value. The result is shown in Fig.~\ref{fig:short-wall-entry-slope}. Firstly, we computed this quantity by numerically solving the full wave equation (\ref{eq:wave}) in the exterior and using the original exterior (Euclidean) integral formula (\ref{eq:euclidean}). That is a purely exterior and exact calculation. We then re-computed the observable using the deformed interior contour, with the wave equation solved in a Born series approximation. This approximation is the same as that described above (\ref{eq:inv}), corresponding to treating the potential in the wave equation (\ref{eq:wave2}) as a perturbation. The Born series is found to converge very quickly. Here we included the first correction to the unperturbed wave (\ref{eq:wkbschw}). These two calculations agree, demonstrating that there is no veil-like obstruction to the deformation in this case. The plot exhibits the anticipated pairs of zeros.

\begin{figure}[h]
  \centering
  \includegraphics[width=0.7\linewidth]{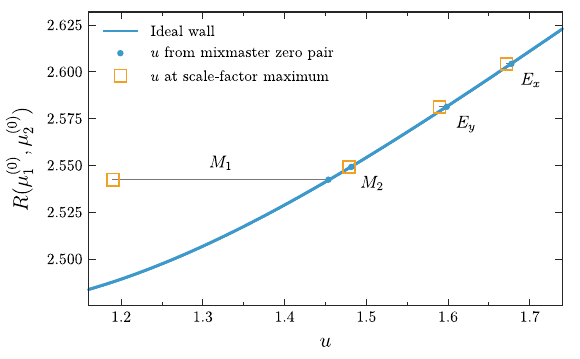}
  \caption{Here $\mu_1^{(0)},\mu_2^{(0)}$ are pairs of zeros extracted from the one-point function in Fig.~\ref{fig:short-wall-entry-slope}. The blue curve is the ideal bounce relation between the ratio $R(\mu_1^{(0)},\mu_2^{(0)})$ of zeros and the bounce parameter $u$, shown in Fig.~\ref{fig:rosen-wall-zero-ratio}. The blue dots show the predicted values of $u$ based on the measured pairs.
   The open squares are the values of $u$ extracted directly from
   the interior mixmaster geometry (lower plot in Fig.~\ref{fig:interior}).   }
  \label{fig:mixmaster-zero-ratio-u}
\end{figure}

In Fig.~\ref{fig:mixmaster-zero-ratio-u} we have used the pairs of zeros in Fig.~\ref{fig:short-wall-entry-slope} together with the universal bounce map of Fig.~\ref{fig:rosen-wall-zero-ratio} to extract the parameter $u$ associated to each bounce.
This boundary thermal one-point function prediction can then be compared with the values of $u$ actually arising in the interior geometry.\footnote{For example, for a magnetic bounce, let $\tau_b$ be the proper time at which $g_{tt}$ reaches its local maximum. We extract $u$ from the transverse metric by
$ u^2=\left.\frac{d\log g_{xx}}{d\log g_{yy}}\right|_{\tau=\tau_b}$. } The
figure shows that while the first bounce is somewhat off, the second, third and fourth bounces are well-captured with increasing accuracy. The conclusion is that the thermal one-point function of a stringy operator, continued to complex scaling dimension as we have done, captures the BKL chaos of the interior geometry.

In the current model, the structure of the Weyl tensor led to a pair of zeros of the one-point function. Those zeros allowed us to accurately reconstruct the parameters of the bounce. More generally, we can similarly solve for the one-point function of a different ideal bounce and hope to extract parameters of the bounce from the `characteristic features' of the one-point function (not necessarily zeros). It would be interesting to explore other models from this perspective.

\section{Conclusions}
\label{sec:discussion}

In this paper, we have provided a string-theoretic anchor for BKL dynamics, by exhibiting so-called mixmaster chaos in the interior of an M-theory AdS$_4$ black hole. We have probed the M-theory background holographically and obtained boundary observables that are sensitive to the interior chaos. The black hole interior manifests itself in analytic continuation of thermal boundary observables as a function of time, space, quantum numbers, or coupling constants. Interior BKL dynamics then implies that, in the presence of boundary sources, such analytic continuation will reveal rich and chaotic features.

A key aspect of BKL chaos are abrupt bounces between Kasner epochs. We have uncovered consequences of these bounces for boundary observables. Our first finding was that bounces produce holographic veils, i.e.~complex co-dimension two extremal area surfaces defined in \S\ref{sec:veil}. We showed that veils hide the real black hole interior from quasi-normal modes, bouncing singularities and high frequency asymptotics of spectral function. Instead these are governed by a magnetic complex singularity. Several questions arise from this observation:
\begin{itemize}
\item Veils are not necessarily the end of the road for thermal two-point function probes of BKL dynamics. A more refined resurgence analysis, as in \cite{Jia:2026ryl,Grassi:2026ekz}, could likely access BKL in the interior. Understanding further nonperturbative features of the thermal two-point function and how they encode BKL dynamics in the presence of veils would be very interesting.

\item An obviously important question is to understand the bulk resolution of BKL singularities. The finite $N$ and stringy effects discussed recently for Schwarzschild-like black holes in
\cite{Komargodski:2026oob, Batra:2026gbq, Buric:2026sit, Dodelson:2026hjm, Liu:2026luf}
do not directly apply to our setup. Firstly, because we have boundary sources turned on. Secondly, because of holographic veiling. It would be very interesting to extend those methods.

\item  Even with the singularity resolved it is possible that holographic veils persist. It would therefore be very interesting to understand their fate in the finite $N$, finite $\lambda$ theory.

\item The asymptotic quasinormal modes are controlled by a complex magnetic singularity. This singularity has more in common with timelike Reissner-Nordstr\"om singularities than with spacelike BKL ones. Timelike singularities are sometimes resolved in string theory through condensation of branes or strings, e.g.~\cite{Strominger:1995cz, Johnson:1999qt, Polchinski:2000uf, Johnson:2001us}. Any such resolution here may lead to interesting features in the QNM spectrum.

\item An asymptotically AdS$_5$ BKL interior was recently found in \cite{Hennigar:2026pmd}. It would be interesting to adapt our probes to that case and study the structure of holographic veils, which will lead to signatures of BKL chaos in ${\mathcal N} = 4$ SYM theory. Other consistent truncations of string theory that are capable of describing mixmaster interiors with an AdS asymptotics are those studied in \cite{Gauntlett:2009dn, An:2022lvo, Gubser:2009qm, Cai:2023igv}.

\end{itemize}

The veils are extremal surfaces and therefore connect naturally to the holographic entanglement entropy. In \S\ref{sec:entanglement} we have shown how analytically
continued holographic entanglement picks out the areas of holographic veils, giving a direct link to BKL bounces. Questions arising from this fact include the following:
\begin{itemize}
\item Entanglement entropy can distinguish a thermofield double state from a pure state of a one-sided black hole with an end-of-the-world brane in the interior \cite{Maldacena:2001kr, Hartman:2013qma,Cooper:2018cmb}.
Our background has a pure state cousin, with an M-theory end-of-the-world brane \cite{Horava:1995qa,Horava:1996ma,Lambert:2015ita}. The pure state has two competing Ryu--Takayanagi surfaces, and hence one-sided entanglement entropy can distinguish it from the thermal state geometry.

\item Does our analytic continuation of $S_\text{EE}({\mathcal R}_\ell)$ have a physical or operational meaning?

\item Are there other, perhaps more direct, boundary information-theoretic observables that capture the interior BKL dynamics? Holographic complexity is a closely related probe of the interior and may well be interesting to consider \cite{Stanford:2014jda}.

\end{itemize}

Finally, we applied the thermal stringy one-point function calculation of \cite{Grinberg:2020fdj} to our mixmaster interior. We computed the thermal expectation value $\langle \ocal \rangle$ arising from a higher derivative coupling $\phi \, W^2$, as a function of an analytically continued dimension $\Delta = \frac{3}{2} - i \mu$. We demonstrated that $\ocal(\mu)$ effectively encodes the black hole interior and Kasner exponents through the location of its zeros. 
Questions arising from this fact include the following:
\begin{itemize}
    \item As we explained above, focusing on a single higher-derivative interaction is likely not a controlled approximation if we are thinking of tuning $\Delta$ by analytically continuing in the string coupling. It is therefore an open question if stringy thermal one-point functions in an actual top-down model like ABJM can be used to effectively probe BKL dynamics.

    \item The expectation value $\langle \ocal \rangle$ has chaotic zeros along a complex line $\Delta = \frac{3}{2} - i \mu$. It is impossible not to be reminded of the Riemann zeta function (and $L$-functions more generally). While the connection here is entirely speculative there are two intriguing facts: \begin{itemize}
    \item With $\mu \in \mathbb{R}$ this scaling dimension defines a principal series representation of the conformal group $SO(3,2)$. This fact underpins the trivial removal the phase of $F(\mu)$ in Fig.~\ref{fig:short-wall-entry-slope}. The zeta function critical axis, and that of other $L$-functions, is also closely related to dilatation symmetries and principal series representations. See e.g.~\cite{Berry1999, Hartnoll:2025hly} for a physical approach.
    
    \item The semiclassical quantisation of BKL chaos leads to wavefunctions that are automorphic forms on the modular billiard domains. The dilatation-basis wavefunctions are the associated automorphic $L$-functions \cite{Hartnoll:2025hly, DeClerck:2025mem}, which have chaotic zeros along a critical axis.
    \end{itemize}

\item Our black branes have intrinsically flat `Bianchi type I' spatial slices, with the bounces coming from matter fields. In 4d vacuum gravity, Bianchi types VIII and IX underpin homogeneous models of BKL dynamics, with spatial curvature driving some of the transitions between Kasner epochs. It would be interesting to investigate how the involvement of spatial curvature in the interior dynamics changes our probes. For example, the Weyl tensor may no longer vanish at the bounces. In that case one would have to work harder to reconstruct the interior from the one-point function. Also, do veils obstruct access to the singularity in all realisations of BKL dynamics?

\item Generic BKL dynamics generates strong spatial inhomogeneity near the singularity. It would be very interesting to extend the one-point function probe to an inhomogeneous setting.

\item We were able to deform the integration contour to the region close to the singularity without obstruction. It would be interesting to understand if this is always the case and, perhaps relatedly, whether the complex magnetic singularity, that controls the two-point functions, has a role to play in the one-point function also.
    
\end{itemize}

In this paper we have seen how analytic continuation of boundary quantities can lead to `clocks' that probe the black hole interior. This is reminiscent of attempts to extend the boundary Wilsonian RG flow through the horizon, where the boundary partition function with a radial cutoff \cite{Heemskerk:2010hk, Faulkner:2010jy, McGough:2016lol, Hartman:2018tkw} becomes the interior wavefunction of the universe \cite{Araujo-Regado:2022gvw, Hartnoll:2022snh, Blacker:2026poc}. Is there a continued observable in the boundary field theory that encodes the standard clocks of bulk quantum cosmology?

\section*{Acknowledgments}

This work has been partially supported by STFC consolidated grant ST/T000694/1. SAH is partially supported by Simons Investigator award \#620869 and by the OpenAI ChatGPT for Academic Researchers program. The code used to compute and characterise the numerical backgrounds has been written by OpenAI’s ChatGPT and Codex agents.

\appendix

\section{Equations of motion, numerics and stability}
\label{ap:equations}

\paragraph{Equations:} The equations of motion for the ansatz (\ref{eq:ansatz}) following from the action (\ref{eq:thy}) are
\begin{align}
\left(\frac{s f}{z^2}B'\right)' & = - \frac{sf}{4} \left(e^{-2 B} w_x^{\prime 2} - e^{2 B} w_y^{\prime 2}\right)  \,, \label{eq:e1} \\
(s f e^{-2B}w_x')' & = \frac{s w_x w_y^2}{4} \,, \\
(s f e^{2B}w_y')' & = \frac{s w_y w_x^2}{4} \,, \\
\frac{s'}{s} & = - z B^{\prime 2} - \frac{z^3}{4} \left(e^{-2B} w_x^{\prime 2} + e^{2 B} w_y^{\prime 2} \right) \,, \label{eq:e4} \\
f' & = \frac{3(f-1)}{z} + z f B^{\prime 2} + \frac{z^3 f}{4} \left(e^{-2B} w_x^{\prime 2} + e^{2 B} w_y^{\prime 2} \right) + \frac{z^3}{16} w_x^2 w_y^2 \,. \label{eq:e5}
\end{align}

The above equations (\ref{eq:e1}) --- (\ref{eq:e5}), together with a regular horizon at which $f=0$, imply two conservation laws
\begin{align}
\frac{sf}{z^2}B'
 +\frac{sf}{4}(e^{-2B}w_xw_x'-e^{2B}w_yw_y') & =0 \,, \\
\frac{sf'+2fs'}{z^2}
 -\frac{sf}{2}(e^{-2B}w_xw_x'+e^{2B}w_yw_y') & =-\frac{4\pi T}{z_\mathcal{H}^2} = - 2 \k^2_4 \, s_\text{BH} T \,. \label{eq:fconst}
\end{align}
These conservation laws follow from scaling symmetries of the action (cf.~\cite{Gubser:2009cg}) and can be used as a check on the numerical accuracy. In the final step in (\ref{eq:fconst}) we introduced the Bekenstein-Hawking entropy $s_\text{BH} \equiv \frac{4 \pi}{2\k^2_4} \frac{1}{z_\mathcal{H}^2}$. This final conservation law has thermodynamic information. Evaluating (\ref{eq:fconst}) at the boundary $z \to 0$, using (\ref{eq:nbdy}) and (\ref{eq:vevs}), one obtains the Smarr formula
\be
s_\text{BH} T = \frac{3 \langle T_{tt} \rangle + W_x \langle J_x \rangle + W_y \langle J_y \rangle}{2} \,.
\ee
This formula in turn allows the free energy density $\mathcal{F}$ of a black hole solution to be evaluated in terms of purely boundary data
\be\label{eq:fdensity}
\mathcal{F} \equiv \langle T_{tt}\rangle - s_\text{BH} T =  
- \frac{\langle T_{tt} \rangle + W_x \langle J_x \rangle + W_y \langle J_y \rangle}{2} 
\,.
\ee

The conservation law (\ref{eq:fconst}) furthermore excludes the possibility of an inner horizon, following the logic in \cite{Hartnoll:2020fhc, Cai:2020wrp}. Eq.~(\ref{eq:fconst}) requires $\frac{sf'}{z^2}$ to be the same on every horizon (where $f=0$). This quantity is negative on the outer horizon (which has $f'<0$ and $s>0$) but would be positive on a first inner horizon (which would have $f'>0$ but still $s>0$, $s$ cannot change sign, from (\ref{eq:e4})).

\paragraph{Numerical methods:} The equations of motion above can be solved numerically using a shooting method. Regular initial conditions are set at the horizon and then integrated outwards. As this is fairly standard we will be brief. Firstly, a scaling symmetry of the equations allows us to put $z_\mathcal{H}=1$. Thus $f(1)=0$ defines the horizon. Scaling symmetries furthermore allow us to set $B(1)=0$ and $s(1)=1$. The two shooting parameters are then $w_x(1)$ and $w_y(1)$. For every choice of these parameters with $w_x(1)^2 w_y(1)^2<48$ there is a unique regular series expansion solution to the equations of motion at the horizon. Using this expansion to set the initial data and then integrating the equations towards the boundary, one obtains the full geometry. This is consistent with the fact that there are two dimensionless parameters that must be fixed at the boundary, as we described in the main text. The fields obtained in this way will have boundary values $B(0) \neq 0$ and $s(0) \neq 1$, in contradiction with (\ref{eq:asymptotic}). To obtain solutions with the correct asymptotic normalisation we must then rescale $B(z) \to B(z) - B(0)$, $s(z) \to s(z)/s(0)$, $w_x(z) \to e^{-B(0)} w_x(z)$ and $w_y(z) \to e^{B(0)} w_y(z)$. Finally, any dimensionless boundary observables will be insensitive to the choice of $z_\mathcal{H} = 1$.

The same series expansion at the horizon can be used to set initial conditions for interior numerical evolution. To maintain accuracy to very large $z$ we work in terms of the logarithmic variable $\log \frac{z}{z_\mathcal{H}}$.

\paragraph{Symmetry breaking at $T=0$:} An exact solution to (\ref{eq:e1}) --- (\ref{eq:e5}) is given by \cite{Devecioglu:2014iia}
\be\label{eq:lif}
f = \frac{3 - \sqrt{6}}{2} \,, \qquad s = \frac{1}{z^{\sqrt{6}}} \,, \qquad B = 0 \,, \qquad w_x = w_y = \frac{24^{1/4}}{z} \,.
\ee
This solution is an example of what is known as a Lifshitz background, in this case with dynamical critical exponent $z_L = 1+\sqrt{6}$. Such geometries commonly arise as the IR, i.e.~$z \to \infty$, of $T=0$ holographic backgrounds \cite{Hartnoll:2018xxg}.
At zero temperature $z \to \infty$ is the singular `horizon' of the exterior, not the black hole singularity.
The solution (\ref{eq:lif}) is symmetric and, indeed, describes the IR of the symmetric phase at $T=0$ \cite{Grandi:2021bsp}.

The instability of the symmetric phase can be seen from the scaling dimensions of perturbations about the Lifshitz background (\ref{eq:lif}). Let $X = \{f,s,B,w_x,w_y\}$ and let $X_o$ be the solutions in (\ref{eq:lif}). We then perturb $X = X_o (1 + \delta X z^\Delta)$, with $B = \delta B z^\Delta$, in equations (\ref{eq:e1}) --- (\ref{eq:e5}). The linear system for the $\delta X$ has solutions for the follow dimensions only
\be
\Delta = 3 + \sqrt{6} \,, \qquad \Delta = \frac{3 + \sqrt{6} \pm i \sqrt{1 + 10 \sqrt{6}}}{2} \,, \qquad \Delta = \frac{3+\sqrt{6} \pm \sqrt{79+22\sqrt{6}}}{2} \,.
\ee
The complex dimensions here indicate an unstable background \cite{Hartnoll:2018xxg}. The instability of the Lifshitz background (\ref{eq:lif}) is already known from \cite{Grandi:2021bsp}.

\paragraph{Symmetry restoration at $T=T_\text{c}$:} Exponentiation of the unstable mode leads to a spontaneous breaking of the rotational symmetry described in the main text. That is, there is a second $T=0$ background with normalisable modes near the AdS boundary that break the symmetry. Heating up to nonzero temperatures introduces a nonzero-size horizon. As the temperature is increased this horizon `eats up' the $T=0$ Lifshitz IR geometry. Heuristically, at the critical temperature $T_\text{c}$ the unstable mode no longer fits in the IR geometry and hence the symmetry breaking background either ceases to exist (at a continuous transition) or becomes energetically disfavoured (at a first order transition). See e.g.~\cite{Hartnoll:2018xxg}. Evaluating the free energy (\ref{eq:fdensity}) on the numerical backgrounds shows that the transition is continuous in this case. Fig.~\ref{fig:phase} shows the current imbalance in the background with lowest free energy for given sources.

\section{Hamiltonian analysis}
\label{ap:hamiltonian}

\paragraph{Derivation of Hamiltonian:} To obtain the effective Hamiltonian for our system, we must first `ungauge' the metric slightly and write (\ref{eq:ansatz}) as
\be
ds^2_4 = \frac{1}{z^2} \left(-s^2 f dt^2 + \frac{dz^2}{f} + e^{2 A} dx^2 + e^{-2 B} dy^2 \right) \,.
\ee
We will set $A=B$ again at a later stage. This metric, and the gauge field in (\ref{eq:ansatz}), can be substituted into the action (\ref{eq:thy}). The action should be integrated by parts to remove all second-derivative terms. Let us define the Lagrangian via $S \equiv \frac{1}{2\k_4^2} \int dz L$, and work directly with $L$. The factors of $\frac{1}{2\k_4^2}$ can easily be re-instated if needed. Then, the momenta for all fields $q$ are defined in the usual way $\pi_q \equiv \frac{\pa L}{\pa q'}$, where prime denotes a derivative with respect to $z$. 

Upon computing the momenta, one finds that they are not all independent but instead related by the primary constraint
\be\label{eq:primary}
\Phi \equiv \pi_s - \frac{2f}{s} \pi_f + \frac{6 f}{z^3}e^{A - B} \approx 0 \,.
\ee
As usual in a Dirac constraint analysis, $\approx$ means that this relation is only imposed on the constraint surface.
The Hamiltonian is $H \equiv \sum_q q' \pi_q -L$, where the sum is over all six fields $\{A,B,w_x,w_y,f,s\}$. 
The Hamiltonian should be written as a function of coordinates and momenta, but the primary constraint (\ref{eq:primary}) prevents us from expressing all of the time derivatives in terms of momenta. To overcome this difficulty we set $s' \equiv \lambda$, a Lagrange multiplier. Thus we obtain
\be\label{eq:Ham}
H = H_o + \lambda \Phi \,,
\ee
where $H_o$ depends on $s$ but not $\pi_s$ or $\lambda$.

A secondary constraint is obtained by requiring that the primary constraint (\ref{eq:primary}) be conserved under evolution by (\ref{eq:Ham}). That is,
\be\label{eq:HamC}
\pa_z \Phi + \{\Phi,H\} \approx 0 \,.
\ee
This new constraint Poisson commutes with $\Phi$ and does not lead to any further constraints. However, we can now impose the gauge $A - B \approx 0$. Requiring that this gauge is consistent with time evolution, $\{A-B,H\} \approx 0$, imposes the final constraint (setting $A=B$)
\be\label{eq:pif}
\pi_f \approx \frac{4s}{z^3} \,.
\ee
The primary constraint (\ref{eq:primary}) then leads to
\be\label{eq:pis}
\pi_s \approx \frac{2 f}{z^3} \,.
\ee

Given all the constraints, we can obtain the reduced Hamiltonian on the constrained surface. This is a reduced phase space. We can use (\ref{eq:HamC}) to eliminate $\pi_A$ together with (\ref{eq:pif}) and (\ref{eq:pis}) to eliminate $s$ and $\pi_s$. In this way we write, up to a total derivative,
\be\label{eq:HHred}
\sum_q \pi_q q' - H \approx \hat \pi_f f' + \hat \pi_B B' + \pi_x w_x' + \pi_y w_y' - H_\text{red} \,.
\ee
The hats on the momenta conjugate to $f$ and $B$ are because these take different expressions on the constrained surface. To see this, note from (\ref{eq:pif}) and (\ref{eq:pis}) that
\be\label{eq:shift}
s' \pi_s + f' \pi_f = f' \frac{\pi_f}{2} + \frac{3 f \pi_f}{2 z} + \left(\frac{f \pi_f}{2}\right)' \,.
\ee
This expression is to be used in (\ref{eq:HHred}). The total derivative term can be dropped as (\ref{eq:HHred}) only needs to hold under an integral over $z$.
The momentum conjugate to $f$ thus becomes
\be
\hat \pi_f = \frac{\pi_f}{2} \,.
\ee
From (\ref{eq:HHred}), the second term in (\ref{eq:shift}) must be subtracted from the effective Hamiltonian
\be
H_\text{red} = H - \frac{3 f \pi_f}{2 z} \,.
\ee
Finally, we must note that imposing $A = B$ gives
\be
A' \pi_A + B' \pi_B = B'(\pi_A + \pi_B) \,.
\ee
This means that in the reduced dynamics the momentum conjugate to $B$ is now
\be
\hat \pi_B = \pi_A + \pi_B \,.
\ee
Putting all of the above together we land on the reduced Hamiltonian
\be
H_\text{red} = \frac{48(f-1) + z^4 w_x^2 w_y^2}{16 z} \,\hat \pi_f - \frac{z^2 \hat \pi_B^2 + 4 \left(e^{2 B}\pi_x^2 + e^{-2 B} \pi_y^2 \right)}{4 z^3 f} \frac{1}{\hat \pi_f}\,.
\ee
This is the result we have quoted in (\ref{eq:Hred}) in the main text.

\paragraph{Canonical transformation:} Introduce $\{u,v,\tau\}$ by
\be
u = \frac{B-B(z_\mathcal{H})}{\log \frac{z}{z_\mathcal{H}}} \,, \qquad v^2 + u^2 + 3 =  - \frac{\log \frac{-f \hat \pi_f^2}{4}}{\log \frac{z}{z_\mathcal{H}}} \,, \qquad e^\tau = 2 v \log \tfrac{z}{z_\mathcal{H}} \,.
\ee
The factors of $z_\mathcal{H}$ and $B(z_\mathcal{H})$ are optional, but set convenient initial conditions for $u,v$ at the horizon (namely $u \to 0$ and $v \to \infty$). Indeed, these factors are removed from the final expression in (\ref{eq:potA}) below by further rescaling
\be
w_x = \frac{e^{B(z_\mathcal{H})}}{z_\mathcal{H}} \hat w_x \,, \quad w_y = \frac{e^{-B(z_\mathcal{H})}}{z_\mathcal{H}} \hat w_y \,, \quad \pi_x = \frac{z_\mathcal{H}}  {e^{B(z_\mathcal{H})}} \hat \pi_x \,, \quad \pi_y = \frac{z_\mathcal{H}}  {e^{-B(z_\mathcal{H})}} \hat \pi_y \,.
\ee
We only perform this rescaling for the purposes of writing down the Hamiltonian $H_\tau$ below.

The remaining variables in the transformation, $\{\pi_u, \pi_v, H_\tau\}$, are then found by requiring
\be
\hat \pi_f df + \hat \pi_B dB - H_\text{red}dz = \pi_u du + \pi_v dv - H_\tau d\tau + 2 d(\hat \pi_f f) \,.
\ee
Reading off the coefficients we obtain
\be
H_\tau^2 = v^2 \left(\pi_u^2 + \pi_v^2 \right) + V_\tau(u,v) \,,
\ee
with the time-dependent potential $V_\tau(u,v)$ given by
\be\label{eq:potA}
V_\tau(u,v) \equiv e^{2 \tau} \left[\hat \pi_x^2 e^{-2(1-u) \frac{e^\tau}{2v}}
+\hat \pi_y^2 e^{-2(1+u) \frac{e^\tau}{2v}}
+ \frac{\hat w_x^2 \hat w_y^2}{4} e^{-(u^2+v^2-1) \frac{e^\tau}{2v}} - 12 e^{-(u^2 + v^2 + 3) \frac{e^\tau}{2v}}\right] \,.
\ee
These are equations (\ref{eq:Htau}) and (\ref{eq:V}) in the main text.

\begin{figure}[h]
    \centering
    \includegraphics[width=0.95\linewidth]{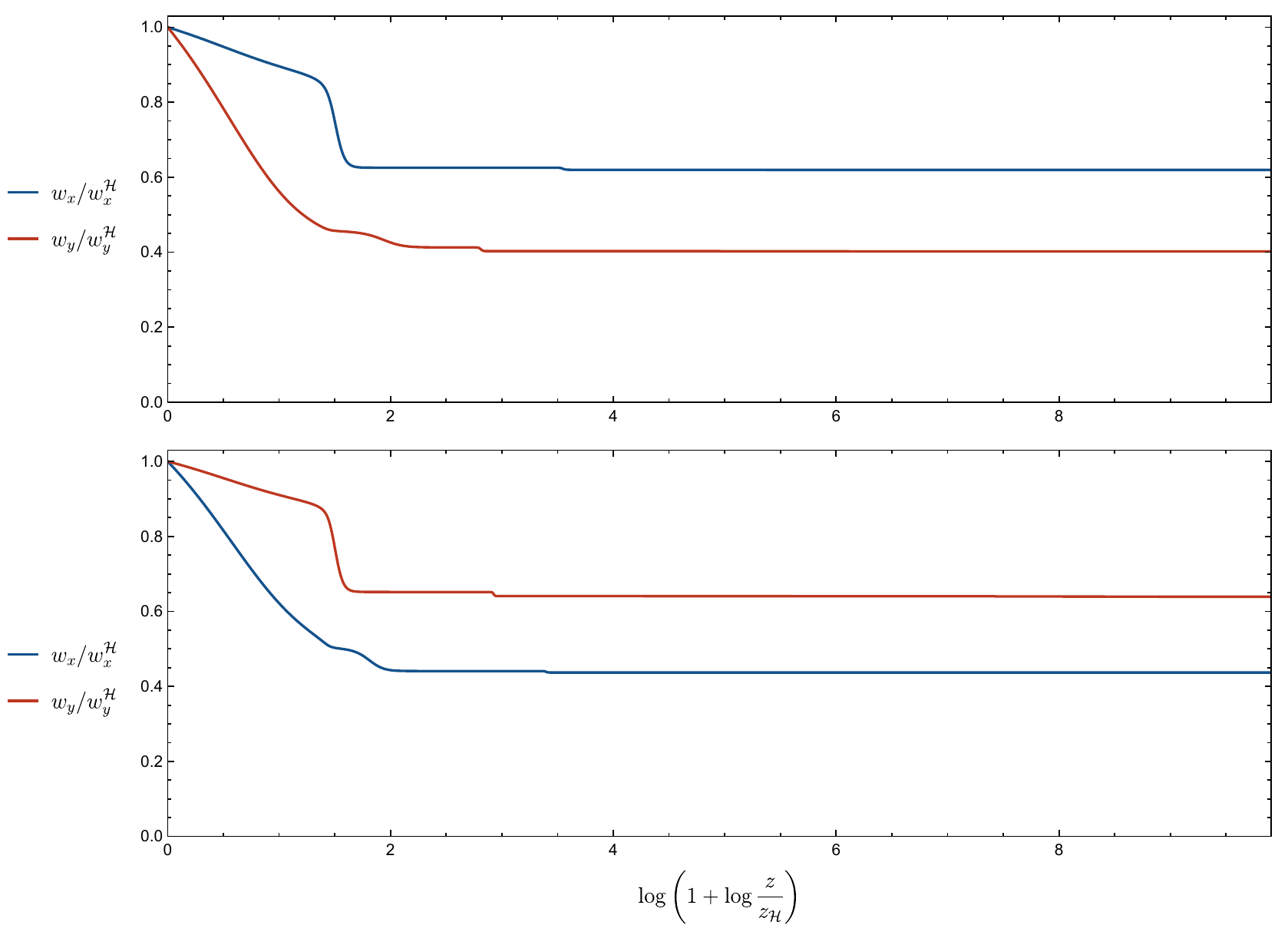}
    \caption{Freezing of bulk gauge fields at late interior times for the interiors shown in Fig.~\ref{fig:interior}. The gauge fields have been normalised by their values on the horizon.}
    \label{fig:gauge}
\end{figure}

\paragraph{Freezing of the gauge fields:} Dropping the final term, which is exponentially small at late interior times, the potential is a sum of three walls
\be
V_\tau = V_x + V_y + V_w \,.
\ee
Hamilton's equations for $\{\hat w_x, \hat w_y,\hat \pi_x, \hat \pi_y\}$ are then
\be\label{eq:hameq}
\dot{\hat{w}}_x = \frac{V_x}{H_\tau \hat \pi_x} \,, \qquad \dot{\hat{w}}_y = \frac{V_y}{H_\tau \hat \pi_y} \,, \qquad \dot{\hat{\pi}}_x = -\frac{V_w}{H_\tau \hat w_x} \,, \qquad \dot{\hat{\pi}}_y = -\frac{V_w}{H_\tau \hat w_y} \,.
\ee
Here $\dot{} \equiv \frac{d}{d\tau}$. Away from the walls the numerators in (\ref{eq:hameq}) are exponentially small. So long as the denominators are nonzero, then, the $\{\hat w_x,\hat w_y,\hat \pi_x,\hat \pi_y\}$ variables do not change appreciably. These variables do change over a bounce off a wall.
However, such changes also become exponentially small at late interior times due to the bounces becoming very sudden. We give an example of this phenomenon in the following paragraph, see also Appendix A of \cite{Damour:2002et}. Fig.~\ref{fig:gauge} illustrates the freezing of the gauge fields using the numerical backgrounds discussed in the main text.

Consider a bounce off the $u=1$ wall. Over the bounce we may keep only the $V_x$ term in the potential, as the others are exponentially small. In this case we have the Hamilton equation
\be
\dot{\pi}_u = - \frac{e^\tau}{2v} \frac{V_x}{H_\tau} \,,
\ee
and hence over a bounce, combining with (\ref{eq:hameq}),
\be\label{eq:Delta}
\Delta \hat w_x = - \frac{2 v_\star e^{-\tau_\star}}{\hat \pi_x} \Delta \pi_u \,.
\ee
Here we used the fact that $\hat \pi_x$ is conserved over this bounce, as $V_w$ is exponentially small in (\ref{eq:hameq}), and that the bounce is fast at late times so that $v$ and $\tau$ have well-defined values $v_\star$ and $\tau_\star$ over the bounce. The momentum $\pi_u$ changes sign, so $\Delta \pi_u = - 2 \pi_u^\text{in}$. The $e^{-\tau_\star}$ in (\ref{eq:Delta}) leads to exponentially small changes in $\hat w_x$ only.

\section{Numerical checks of quasinormal modes}
\label{app:qnm}

In this appendix we perform high-resolution checks on the analytic formula (\ref{eq:analyt}) for the asymptotic QNMs, using the specific interior shown in the lower plot of Fig.~\ref{fig:interior}.
We must first give the initial conditions at the horizon to high accuracy. As explained in Appendix \ref{ap:equations} above, the numerics can always be performed setting the horizon to $z_{\mathcal{H}}=1$ together with $f(1)=0$, $B(1)=0$ and $s(1)=1$. For the particular background we made the choice of $w_x(1) = 0.9306048591020996$ and $w_y(1) = 1.8612097182041991$. As also explained in Appendix \ref{ap:equations}, to convert to the physical boundary normalisation we use the boundary values that follow after solving the equations. In this case: $s(0) \approx 1.0118289139129633$ and $B(0) \approx -0.001522537870944675$.

Using the background solution with the properly boundary-normalised fields we find that the optical depth $\mathcal{Z}_\text{QNM}\approx 0.50956894+1.11948844 i$ while the constants in (\ref{eq:Vexp}) are 
$c_1 \approx 1.67077-2.20199 i$,
$c_2\approx -0.21886-8.53174 i$ and $c_3 \approx -10.44663-20.55755 i$. The temperature, still in units where $z_{\mathcal{H}} = 1$, is $T \approx 0.22119514$. To remove the dependence on this choice of units, we must consider dimensionless quantities such as $\omega_n/T$. We compute the QNM frequencies following the method described in \cite{Hartnoll:2026vhu}. A selection of numerically computed frequencies are shown in Table \ref{tab:QNM}.

\begin{table}[h]
\centering
\begin{minipage}{0.4\linewidth}
\footnotesize
 \setlength{\tabcolsep}{4pt}
\begin{tabular}{ccc}
\toprule
$n$ & $\displaystyle \operatorname{Re} \frac{\omega_n}{T}$ & $\displaystyle \operatorname{Im} \frac{\omega_n}{T}$\\
\midrule
0  & 1.536865 & $-4.535782$ \\
1  & 5.595929 & $-12.16749$ \\
5  & 26.73246 & $-53.11050$ \\
10 & 51.26243 & $-104.8282$ \\
20 & 99.48355 & $-208.7151$ \\
40 & 194.9458 & $-417.2907$ \\
80 & 384.8294 & $-836.3932$ \\
\bottomrule
\end{tabular}
\end{minipage}
\hspace{0.8cm}
\begin{minipage}{0.25\linewidth}
\footnotesize
 \setlength{\tabcolsep}{4pt}
 \begin{tabular}{ccc}
 \toprule
 $j$ & $\frac{\gamma_j}{T}$
     & $\frac{\gamma_j-\gamma_{j-1}}{T}$ \\
 \midrule
 1 & 1148.998 & --- \\ 
 2 & 1782.645 & 633.646 \\ 
 3 & 2425.500 & 642.855 \\ 
 4 & 3073.741 & 648.241 \\ 
 5 & 3725.300 & 651.559 \\ 
 6 & 4379.041 & 653.74 \\
 \bottomrule
 \end{tabular}
\end{minipage}
\caption{Left: Real and imaginary parts of selected QNM frequencies, computed numerically from the exterior of the lower background in Fig.~\ref{fig:interior}. Right: first few purely imaginary modes in the same background.
\label{tab:QNM}}
\end{table}

\begin{figure}[h]
    \centering
    \includegraphics[width=0.8\linewidth]{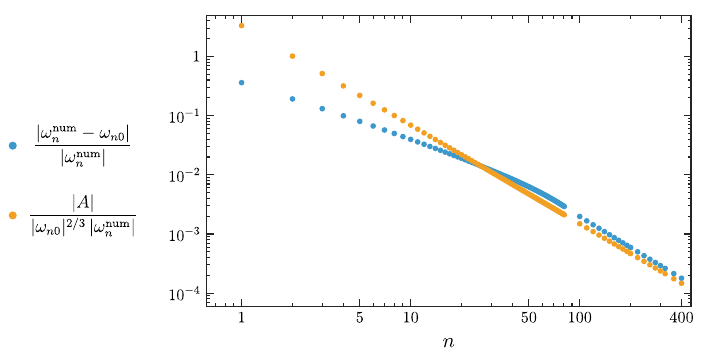}
    \caption{Blue dots: relative difference between the numerical QNM $\omega_n^\text{num}$ and the leading order WKB prediction $\omega_{n0}$ in (\ref{eq:wn0}). Orange dots are the prediction for this difference from the WKB correction (\ref{eq:analyt}), with $A$ the coefficient in (\ref{eq:analyt}).}
    \label{fig:comp}
\end{figure}

In Fig.~\ref{fig:comp} we compare the WKB prediction (\ref{eq:analyt}) for this same background, which comes from the complex magnetic singularity (\ref{eq:magsing}), with the numerical QNM frequencies. The comparison shows that the leading $\omega_{n0}^{-2/3}$ power-law correction and the magnitude of its coefficient in (\ref{eq:analyt}) are consistent with the numerical results. In particular, the difference between the two curves in the plot is consistent with subleading $\frac{\log \omega_{n0}}{\omega_{n0}}$ and $\frac{1}{\omega_{n0}}$ corrections, but $n$ is not large enough to pin down the coefficient of these terms.

The lowest overdamped modes found numerically are shown in Table \ref{tab:QNM}. The table also shows the difference between sequential modes. This spacing is seen to be tending towards the asymptotic prediction from (\ref{eq:gamma}) that
\be
\frac{\Delta \gamma_\infty}{T} = {\pi \over {T \mathcal Z}_{\text{I}}} \approx 661.5 \,,
\label{eq:predasym}
\ee
with $\mathcal{Z}_I$ obtained for the background using (\ref{eq:ZZ}).

\section{Leading asymptotics of the WKB solution}
\label{app:normone}

Here we prove in generality the statement (\ref{eq:wkbschw}) in the main text that at large $\rho$
\be\label{eq:obj}
\psi_{\ocal}(\rho)= - e^{i \mu \ell_{\text{hor}}} {e^{- i \mu \rho} \over \sqrt{v}} \,, 
\ee
to leading order in the large $\mu$ WKB expansion.

Close to the horizon, $v\sim v_1\rho$. The wave equation (\ref{eq:wave2}) thus becomes, at large $\mu$,
\be\label{eq:nh}
G_\ocal''+\frac1\rho G_\ocal'+\mu^2G_\ocal=0 \,.
\ee
The regular solution in this regime is the Bessel function
\be\label{eq:go}
G_{\mathcal O}(\rho) =  g_o  J_0(\mu\rho) \,.
\ee
Now $\psi_\ocal$ also obeys (\ref{eq:nh}) near the horizon, but is required to have purely $e^{-i\mu\rho}$ behaviour at large $\rho$. The near-horizon solution that matches onto this is
\be\label{eq:ppp}
\psi_\ocal(\rho) =  p_o H_0^{(2)}(\mu\rho) \,.
\ee

The Hankel continuation formula is
\be
H_0^{(2)}(e^{-i\pi}x)
=H_0^{(2)}(x)-2J_0(x) \,.
\ee
Now recalling that (\ref{eq:disc}) requires, in the near horizon regime,
\be
\psi_\ocal(e^{-i\pi} \rho)-\psi_\ocal(\rho) = G_\ocal(\rho) \,,
\ee
we see that the monodromy fixes the coefficients
\be\label{eq:pg}
-2 p_o =g_o \,.
\ee
Now we write (\ref{eq:go}) as
\be\label{eq:gnh}
G_{\mathcal O}(\rho) = 
\frac{g_o}{2}\left(H_0^{(1)}(\mu \rho)+H_0^{(2)}(\mu \rho)\right) \,.
\ee
Recall that $G_\ocal(\rho)$ is normalised to have unit source at the boundary. The source term comes from 
the $H_0^{(2)}(\mu \rho)$ near-horizon term in (\ref{eq:gnh}).
From (\ref{eq:ppp}), (\ref{eq:pg}) and (\ref{eq:gnh}), therefore, we see that the source term for $\psi_\ocal$ will be the negative of that for $G_\ocal$.

Recalling that $\rho \approx -\log z+\ell_{\mathrm{hor}}$ as $z \to 0$, the large $\rho$ WKB solution for $G_\ocal(\rho)$ is $G_\ocal(\rho) \approx e^{i \mu \ell_{\text{hor}}} e^{-i\mu\rho}/\sqrt v + \cdots$. Changing the sign, as we have just explained, we obtain (\ref{eq:obj}) for $\psi_\ocal(\rho)$.

\bibliographystyle{JHEP}
\bibliography{references}
\end{document}